\documentclass[prodmode,acmtaco]{acmsmall}

\def\finish{\mbox{\tt finish}}

\def\newfinishFull{\mbox{Aggressive Finish-Elimination}}
\def\newfinish{\mbox{AFE}}

\def\dynamicChunkingFull{{Dynamic Load-Balanced loop Chunking}}
\def\dynamicChunking{\mbox{DLBC}}

\def\dcafe{\mbox{DCAFE}}
\def\lc{\mbox{LC}}
\def\UnOpt{\mbox{UnOpt}}

\makeatletter
\newif\if@restonecol
\makeatother

\makeatletter
\def\hlinewd#1{%
\noalign{\ifnum0=`}\fi\hrule \@height #1 %
\futurelet\reserved@a\@xhline}
\makeatother

\usepackage{url}
\usepackage{times}

\usepackage{stmaryrd}
\usepackage{colortbl}
\usepackage{graphicx,color}
\usepackage{caption}

\usepackage[labelformat=simple]{subcaption}
\usepackage[linesnumbered]{algorithm2e}
\usepackage{float}
\usepackage{multirow}
\usepackage{listings}		
\lstdefinestyle{rm}{mathescape,basicstyle=\small\ttfamily}

\DeclareCaptionType{copyrightbox} 
\usepackage{array}

\newcommand{\mc}[1]{\ensuremath{\mbox{\em #1}}} 
\newcommand{\ifTR}[1]{} 
 
\usepackage{code}

\newcommand{\subparagraph}{}

\renewcommand{\dbltextfloatsep}{4pt}
\renewcommand{\textfloatsep}{4pt}
\renewcommand{\floatsep}{2pt}
\renewcommand{\dblfloatsep}{2pt}
\begin{document}

\markboth{S. Gupta et al.}{DCAFE: Dynamic load-balanced loop Chunking \& Aggressive Finish Elimination for 
RTP Programs~}

\title{DCAFE: Dynamic load-balanced loop Chunking \& Aggressive Finish Elimination for 
Recursive Task Parallel Programs}
\author{Suyash Gupta
\affil{Indian Institute of Technology Madras}
Rahul Shrivastava
\affil{Indian Institute of Technology Madras}
V. Krishna Nandivada
\affil{Indian Institute of Technology Madras}}

\begin{abstract}
\label{s:abstract}

In this paper, we present two symbiotic optimizations to optimize recursive task parallel (RTP) 
programs by reducing the task creation and termination overheads.
Our first optimization 
\newfinishFull{}~(\newfinish{}) helps 
reduce the redundant join operations to a large extent.
The second optimization 
\dynamicChunkingFull{} (\dynamicChunking{}) 
extends the prior work on loop chunking to decide on the 
number of parallel tasks based on the number of available worker threads, at runtime. 
Further, we discuss the impact of exceptions on our optimizations
and extend them to handle RTP programs that may throw
exceptions.
We implemented \dcafe{} 
(= \dynamicChunking{}+\newfinish{}) in the X10v2.3 compiler and
tested it over a set of benchmark kernels on two different
hardwares
(a 16-core Intel system and a 64-core AMD system).
With respect to the base X10 compiler extended with loop-chunking of
Nandivada et al~\cite{nandivada13-toplas} (LC),
\dcafe{} achieved a
geometric mean speed up of 5.75$\times$ and 4.16$\times$ on the Intel and
AMD system, respectively.
We also present an evaluation with respect to the energy consumption on
the Intel system and
show that on average,
compared to the \lc{} versions,
the \dcafe{} versions 
consume 71.2\% less energy.

\end{abstract}




\acmformat{Suyash Gupta, Rahul Shrivastava and V. Krishna Nandivada, 2015. 
DCAFE: Dynamic load-balanced loop Chunking \& Aggressive Finish Elimination for 
Recursive Task Parallel Programs.}

\begin{bottomstuff}
New paper, Not an extension of a conference paper.\\
This research is partially supported by the New Faculty Seed Grant, funded
by IIT Madras CSE/11-12/567/NFSC/NANV, DAE research grant CSE/13-14/139/BRNS/NANV and
DST Fasttrack grant CSE/13-14/140/DSTX/NANV.\\
Author's addresses: Suyash Gupta, Rahul Shrivastava {and} V. Krishna Nandivada, 
Department of Computer Science and Engineering,
Indian Institute of Technology Madras, Chennai - 600036. 
\end{bottomstuff}

\maketitle

\section{Introduction}
\label{s:intro}
The onset of multi-core architectures has brought forth a shift in programming 
paradigm from sequential programs to parallel programs.
This shift has led to an increased interest in task parallel 
languages, such as X10~\cite{x10manual}, Chapel~\cite{chapel}, OpenMP~\cite{openMP},
HJ~\cite{hjava}, and so on.
These languages allow the programmer to express the desired amount of 
parallelism (a.k.a {\em ideal} parallelism), while delegating the 
task of extracting the {\em useful} parallelism to the compiler
(or runtime). 
In this paper, 
we present two novel compiler optimizations (targeting recursive task parallel
programs) to extract useful parallelism from the ideal.

Recursive Task Parallel (RTP) programs constitute an important subset of 
task parallel programs.
In RTP programs the parent task spawns a set of child tasks, which in turn can 
recursively spawn further tasks. 
This renders the problem of extracting useful parallelism from the ideal quite
challenging in case of RTP programs (compared to non-RTP
programs).
We will use an example to illustrate the same.

\begin{figure}[t]
\small
\begin{tightcode}
1 def \textbf{find\_queens()} \{
2  ...
3  \textbf{nqueens} (n, 0, ...); \} 
 
4 def \textbf{nqueens}(val n:Int, val j:Int, ...) \{ 
5   \textbf{finish} \{ 
6      for(var i:Int=0; i<n; i++) \{
7         \textbf{async} \{ 
8            ...    /* {\em Checking if none of the queens conflict}  */
9            \textbf{nqueens}(n, j+1, ...); \} \} \} \}
                     (a)
4 def \textbf{nqueens}(val n:Int, val j:Int, ...) \{
5   var \textbf{nChunks}:Int=\textbf{Runtime.retNthreads()};
6   var \textbf{chunkSize}:Int=(n+nChunks-1)/nChunks;
7   \textbf{finish} \{
8      for(var ii:Int=0; ii<n; ii+=chunkSize) \{
9         val ni = ii;
10        \textbf{async} \{ 
11           var kx:Int = ni+chunkSize;
12           if(kx>n) kx=n;
13           for(var i:Int=ni; i<kx; i++) \{ 
14              ...   /* {\em Checking if none of the queens conflict}  */
15              \textbf{nqueens}(n, j+1, ...); \} \} \} \} \} 
                     (b)
\end{tightcode}
\caption{BOTS Nqueens kernel in X10: (a) Unoptimized version (b) Loop Chunked version of the nqueens function.}
\label{fig:motivate-nqueens}
\end{figure}

Figure~\ref{fig:motivate-nqueens}(a) shows the snippet of 
the BOTS~\cite{bots} Nqueens kernel, in X10.
The {\tt async} construct spawns a new child task to execute 
the statement within its body, in parallel with the parent task. 
The {\tt finish} construct acts as a join point for all the tasks spawned 
in its body.
The code in Figure~\ref{fig:motivate-nqueens}(a)
shows that the presence of recursive task parallelism may lead to
the execution of a large number of {\tt finish} operations at runtime (for
example, when n=14, it executes 27 million {\tt finish} operations). 
Prior work~\cite{nandivada13-toplas} shows that eliminating unnecessary {\tt finish} operations
can lead to significant performance improvements. 
However, their proposed technique does not lead to any reduction in the number of 
{\tt finish} operations, in this example.
Interestingly, we observe that
each task spawns new child tasks, and waits at the join point for the 
spawned tasks to terminate. 
After that the task simply returns from the procedure.
Hence, this {\tt finish} construct can be pulled out of the {\tt nqueens} method and placed 
around its non-recursive call site (in {\tt find\_queens}).
Or in other words, 
the {\tt finish} construct, in the method {\tt nqueens} can be declared redundant (and hence removed), if we surround
the non-recursive call to {\tt nqueens} with a {\tt finish} construct. 
Such an optimization helps us in reducing the number of {\tt finish} operations 
to just one (for this code), which can lead to significant performance gains.

In general, it is not trivial to pull out {\tt finish} constructs, as
they may be nested deep inside some {\tt if/while} 
constructs. 
The problem becomes further challenging, if the input code may throw exceptions.
We address these challenges in the first optimization we propose in this paper
(called {\em Aggressive Finish-Elimination}, or {\em \newfinish{}} in short).
\newfinish{} helps to eliminate redundant {\tt finish} operations in RTP programs, in a
semantics preserving way.

Further analysis of 
Figure~\ref{fig:motivate-nqueens}(a) shows that
at the $k^{th}$ level of recursion, the nqueens function creates {\tt n}$^{k}$ number of {\tt async}s
(tasks) leading to
an explosion of tasks (for example, when {\tt n}=14, it creates a total of 377
million tasks), which in turn results in large performance overheads. 
The powerful scheme of Loop Chunking~\cite{nandivada13-toplas} (henceforth referred to as \lc{}) helps to 
extract useful parallelism from the ideal.
\lc{} splits the iterations of a large parallel loop into a set of chunks, where each 
chunk (containing a set of serial iterations) executes in parallel.

Figure~\ref{fig:motivate-nqueens}(b) presents the \lc{} version of the
{\tt nqueens} function.
Here, the call to the {\tt Runtime.retNthreads} function
returns the initial count of the worker threads.
Hence, the useful parallelism is bound by {\tt nChunks}.
Considering this, \lc{} ensures that
at most {\tt nChunks} number of tasks are created in any invocation of this function.
Thus, at level $k$ of recursion, it creates {\tt nChunks}$^k$ number of
tasks (for example,
when {\tt n}=14 and {\tt nChunks}=8, it creates 189 million tasks).
This chunked program runs faster than the unoptimized version, but still
incurs a large task creation and termination overhead. 
This is because the chunking algorithm is oblivious to the recursive call inside the 
loop, and hence, permits the spawning of a large number of tasks.
We have observed a similar trend in a number of RTP kernels present in 
two open-sourced benchmark suites: IMSuite~\cite{imsuite13-arxiv} 
and BOTS.

The main reason for the \lc{} version to incur the large overheads is that it
does not exploit the underlying recursive nature of the task parallel program
and misses significant opportunities to optimize such programs.
To address this challenge,
we propose our second optimization ``\dynamicChunkingFull{}''
(\dynamicChunking{}), as an extension to \lc{}.
\dynamicChunking{} generates
code that spawns new tasks (to execute some iterations of a loop) only if
``idle" workers are available, at runtime. 
Otherwise, the current worker executes the loop serially.
During the serial execution, if some workers get freed up, the remaining loop iterations 
may be executed in parallel (by the available workers).
Our transformation leads to significant reduction in the number of tasks
created:
for example, for Figure~\ref{fig:motivate-nqueens}(a), when n=14, our {\em transformed} code 
creates 6 million tasks ($\approx$ 30$\times$ less, compared to the
\lc{} version).

Realising the above mentioned extensions requires multiple design choices
(e.g., how to identify the number of available workers, how to divide work
among the current and available workers, when to execute the code in
the serial mode, when to switch back to parallel execution, and so on),
that are non-trivial in nature.
We studied many different design alternatives and designed DLBC using the
best available choices.



\noindent 
{\bf Our Contributions }\\
\noindent
$\bullet$ We propose  two 
symbiotic optimizations
\newfinish{} and \dynamicChunking{},
for improving the performance of RTP
programs that reduce the redundant join and task creation operations. 
\dcafe{} (= \dynamicChunking{} + \newfinish{}) can be easily extended to other task parallel languages (such as 
HJ, Chapel and OpenMP) that have similar constructs for task creation and 
task termination operations.
\\
\noindent
$\bullet$
We present an extension to the X10v2.3 compiler that implements \dcafe{}. 
\\
\noindent
$\bullet$
We extend \dcafe{} to perform semantics preserving code 
transformation even in the presence of exceptions. \\
\noindent
$\bullet$
We 
evaluated \dcafe{} 
over 8 benchmarks (drawn from two benchmark suites: IMSuite
and BOTS)  
on two different hardware systems (a 16-core Intel 
system and a 64-core AMD system).
We show that \dcafe{} leads to improved execution times (geometric mean 
of 5.75$\times$ on the Intel and 4.16$\times$ on the AMD system, with respect to
the \lc{} version; and 
geometric mean of 12.64$\times$ on the Intel and 5.25$\times$ on the AMD
system, with 
respect to the unoptimized version).\\
\noindent
$\bullet$
We also show that the use of \dcafe{} leads to significantly lower energy
consumption, on the Intel system.
On average, \dcafe{} optimized codes consume 0.288$\times$ the energy consumed by \lc{} 
optimized codes, and 0.19$\times$ the energy consumed by the unoptimized 
codes. 

{\em Organization}: Section~\ref{s:back} presents a brief background of
some of the topics pertinent to this paper. 
Section~\ref{s:trans} discusses the details of \newfinish{} and
\dynamicChunking{}.
In Section~\ref{s:exception}, we present relevant changes to these
transformations in the presence of exceptions.
In Section~\ref{s:eval}, we present an evaluation of \dcafe{}.
We present a discussion about some of the salient aspects of our work in
Section~\ref{s:analysis} and present a discussion on the related work in
Section~\ref{s:related}.
Finally, we conclude in Section~\ref{s:concl}.

\section{Background}
\label{s:back}
\subsection{X10}
We briefly describe the X10 constructs 
relevant to this manuscript
(see the language manual~\cite{x10manual} for details).
%
%
%
``{\tt async S1}" spawns a new asynchronous task to execute {\tt S1}. 
A task can be registered on one or more clocks. 
``{\tt async clocked(c1,c2) S}" registers the new spawned task on the clocks
{\tt c1} and {\tt c2}.
Such a task executing  {\tt Clock.advanceAll()}, waits for all the tasks
registered on {\tt c1} and/or {\tt c2} to execute the barrier {\tt
Clock.advanceAll()}.
``{\tt finish S1}" waits for all the tasks spawned in {\tt S1} to terminate.
In X10, each {\tt async} has a unique Immediately Enclosing Finish (IEF), 
at runtime. Note: statically an {\tt async}  may have multiple IEFs.

During execution, when an exception is thrown in an 
{\tt async},
it is caught by its IEF.
The enclosing {\tt finish} waits for termination of the remaining tasks, and then
packages all the thrown exceptions as a {\tt MultipleExceptions} object,
and throws it again.
Note: an exception that occurs in one 
task ({\tt async}) does not terminate 
the sibling tasks.

X10 runtime is built around the notion of {\em workers}. 
Each worker is assigned a task to execute and 
can be seen as a software thread.
The initial count for workers can be set (typically to to the number of available cores) at runtime, 
using the environment variable {\tt X10\_NTHREADS}.
During execution, X10 runtime also tracks the number of {\em idle-workers} -- workers 
which are assigned no task.

\subsection{Finish Elimination}
Finish Elimination~\cite{nandivada13-toplas} helps to 
remove redundant {\tt finish} constructs -- {\tt finish} 
constructs that do not contain {\em e-asyncs}.
For a statement {\tt S},
all the {\tt async} statements (within {\tt S}) whose IEF is
not enclosed within {\tt S} are called escaping asyncs or {\em e-asyncs}~\cite{GuoBarikRamanSarkar09} of {\tt
S}.
The `Finish Elimination' optimization
repeatedly applies a series of transformations 
to eliminate the redundant {\tt finish} constructs. 
Three of their proposed set of mini-transformations ({\em Loop-Finish Interchange}, {\em Finish Fusion}
and a simplified version of {\em Tail Finish Elimination}) are relevant to this work. 
For the sake of completeness, we reproduce these rules in
Figure~\ref{fig:existing-optim}.
Each transformation may include a set of pre-conditions (shown as comments) necessary
to ensure semantics preserving transformation.
{\em Loop-Finish Interchange} is feasible when, neither there is a 
loop carried dependence between the iterations of the loop, nor 
the loop condition depends on the e-asyncs of {\tt S3}.
This rule can be trivially extended for other 
looping constructs such as, {\em while} and {\em do-while}.
{\em Finish Fusion} merges two {\tt finish} statements,
if {\tt S2} has no dependence on 
the e-asyncs of {\tt S1}. 
{\em Tail Finish Elimination} eliminates the trivially redundant {\tt finish} 
constructs.

\begin{figure}[t]
\centering
\small
\begin{tabular}{|lll|} \hline

\multicolumn{3}{|l|}{{\bf 1. Loop-Finish Interchange}} \vspace*{-0.05in}\\
\begin{lstlisting}[style=rm]
for(S1;c;s2) { $\mbox{\textbf{finish}}$ S3 }

\end{lstlisting}
& \hspace{-0.2cm}
$\Longrightarrow$
& \hspace{-0.2cm}
\begin{lstlisting}[style=rm]
S1; $\mbox{\textbf{finish}}$ { for(;c;S2) {S3}}
\end{lstlisting} \\
\multicolumn{3}{|l|}{
// {\em Say $E_s$ = set of e-asyncs in {\tt S3}}} \\
\multicolumn{3}{|l|}{// {\em $\lnot \exists e$ $\in$ $E_s$: {\tt c} has dependence on $e$.}} \\
\multicolumn{3}{|l|}{
// {\em $\lnot \exists e$ $\in$ $E_s$: $e$ has loop carried dependence on
{\tt S2}, {\tt c} or {\tt S3}~ ~~~}}\\\hline
\multicolumn{3}{|l|}{{\bf 2. Finish Fusion}} \vspace*{-0.05in}\\
\begin{lstlisting}[style=rm]
$\mbox{\textbf{finish}}${S1} $\mbox{\textbf{finish}}${S2}
\end{lstlisting}
& \hspace{-0.2cm}
$\Longrightarrow$
& \hspace{-0.2cm}
\begin{lstlisting}[style=rm]
$\mbox{\textbf{finish}}${S1; S2}
\end{lstlisting} \\
\multicolumn{3}{|l|}{// {\em {\tt S2} has no dependence on any e-async of {\tt S1}.}} 
\\ \hline
\multicolumn{3}{|l|}{{\bf 3. Tail Finish Elimination (Simplified)}} \vspace*{-0.05in}\\
\begin{lstlisting}[style=rm]
$\mbox{\textbf{finish}}$ $\mbox{\textbf{finish}}$ S1
\end{lstlisting}
& \hspace{-0.2cm}
$\Longrightarrow$
& \hspace{-0.2cm}
\begin{lstlisting}[style=rm]
$\mbox{\textbf{finish}}$ S1
\end{lstlisting}
\\ \hline
\end{tabular}
\caption{Existing mini-transformations.}
\label{fig:existing-optim}
\end{figure}

\subsection{Energy Measurements (Intel specific)}
{\em Running Average Power Limit}  (RAPL)~\cite{rapl} is an interface that exposes the 
Machine Specific Registers (MSRs) to the user application.
MSRs facilitate the measurement of the energy consumed by different
components of the CPU.
The MPES  register (MSR\_PP0\_ENERGY\_STATUS) in MSRs
stores the total energy consumed by all the cores in a node.
We have implemented a function {\tt read\_msr()} to read this register, in
our generated code.
We couldn't find a similar interface for our AMD system.

{\section{Transformation Scheme}
\label{s:trans}
In this section, we discuss 
two novel optimizations:
\newfinishFull{} (\newfinish{}) and \dynamicChunkingFull{} (\dynamicChunking{}).
We propose a new compiler optimization phase called \dcafe{} (=
\dynamicChunking{} + \newfinish{}) that
 combines these two optimizations.
\dcafe{} (overall block diagram shown in Figure~\ref{fig:finReductor}) starts by performing a simple 
{\em may-happen-in-parallel (MHP)} dependence analysis. 
For this work, we 
perform an inter-procedural {\em MHP} analysis, as an extension to that of 
Agrawal et al.~\cite{agarwal-dependence}.
The MHP analysis is used to compute the {\em may-happen-before 
dependence} (MHBD)~\cite{nandivada13-toplas} 
information.
After the MHP analysis, \dcafe{} invokes \newfinish{} and
\dynamicChunking{} optimizations, before doing the code generation.
For the sake of simplicity, in this section, we assume that the programs do not 
throw exceptions.
In Section~\ref{s:exception}, we 
extend our proposed optimizations to do semantics preserving 
transformations of X10 programs that may throw exceptions.

\begin{figure}
\centering
\includegraphics[width=0.6\textwidth]{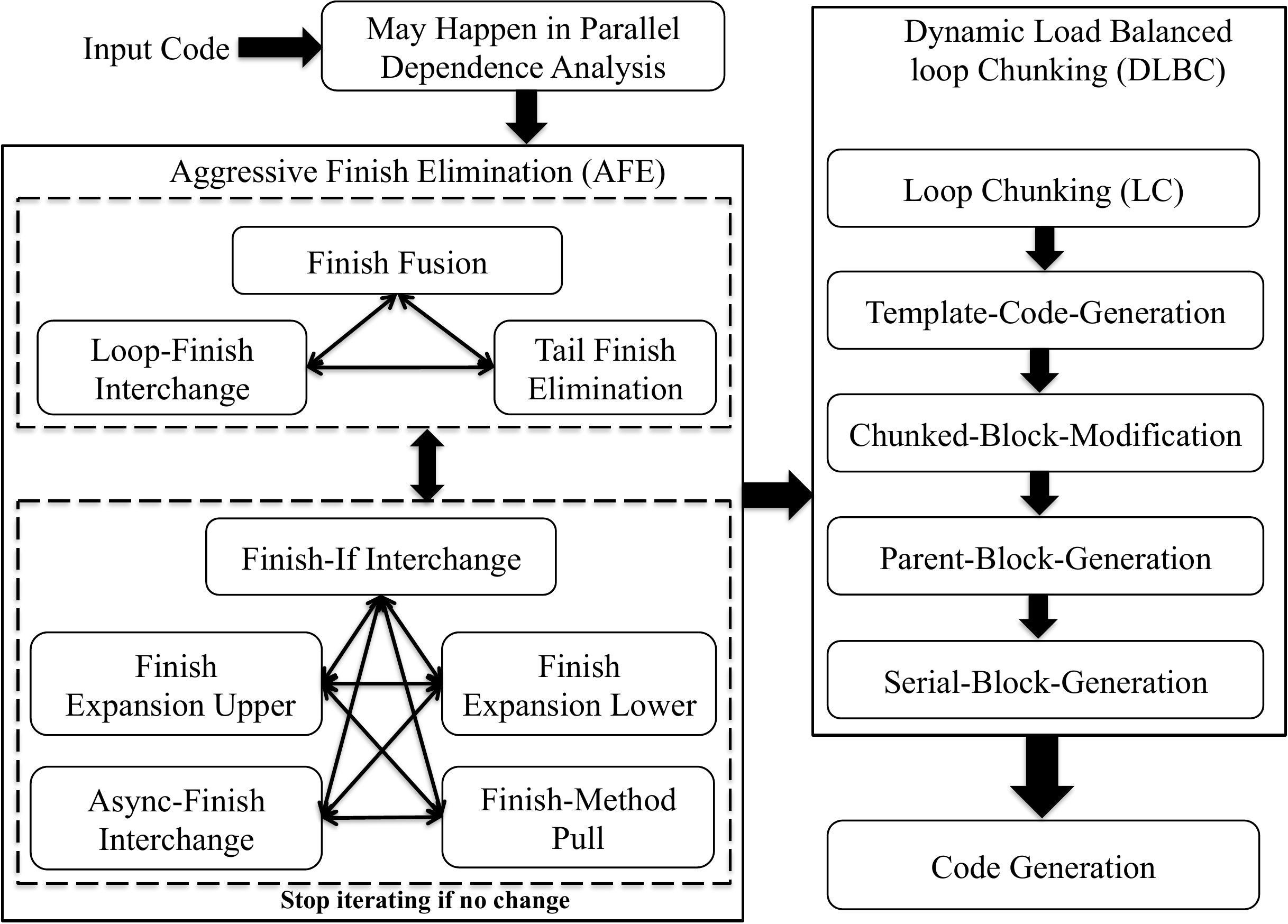}
\caption{Block diagram of \dcafe{}}
\label{fig:finReductor}
\end{figure}

\subsection{\newfinishFull{} (\newfinish{})}
%
\newfinish{} aims at elimination of redundant {\tt finish} 
constructs, and expanding the scope of {\tt finish} operations, if possible. 
\newfinish{} consists of 
eight mini-transformations that aim to 
pull out {\tt finish} constructs from different methods to their respective call-sites.
Three of these mini-transformations
have been proposed by Nandivada et al~\cite{nandivada13-toplas} (Figure~\ref{fig:existing-optim}).
The rest five transformations ({\em Async-Finish Interchange}, 
{\em Finish-If Interchange}, {\em Finish Expansion Upper}, {\em Finish Expansion Lower},  
and {\em Finish-Method Pull}), shown in 
Figure~\ref{fig:new-finish-opt}, are new.
The necessary pre-conditions for any mini-transformation are specified as comments.

\begin{figure}[t]
\centering
\small
\begin{tabular}{|lll|}\hline
\multicolumn{3}{|l|}{{\bf 1. Finish-If Interchange}} \\
\begin{lstlisting}[style=rm]
if(e) {$\mbox{\textbf{finish}}$ S1}
\end{lstlisting}
& \hspace{-0.4cm}
$\Longrightarrow$
&
\begin{lstlisting}[style=rm]
v=e; $\mbox{\textbf{finish}}$ {if(v) S1}
\end{lstlisting}
\\ \hline
\multicolumn{3}{|l|}{{\bf 2. Finish Expansion Upper}} \\
\begin{lstlisting}[style=rm]
S1; $\mbox{\textbf{finish}}$ {S2}
\end{lstlisting}
& \hspace{-0.4cm}
$\Longrightarrow$
&
\begin{lstlisting}[style=rm]
$\mbox{\textbf{finish}}$ {S1; S2}
\end{lstlisting} \\
\multicolumn{3}{|l|}{// {\em If {\tt S1} has no e-asyncs registered on clocks.}}
\\ \hline
\multicolumn{3}{|l|}{{\bf 3. Finish Expansion Lower}} \\
\begin{lstlisting}[style=rm]
$\mbox{\textbf{finish}}$ {S1}; S2
\end{lstlisting}
& \hspace{-0.4cm}
$\Longrightarrow$
&
\begin{lstlisting}[style=rm]
$\mbox{\textbf{finish}}$ {S1; S2}
\end{lstlisting} \\
\multicolumn{3}{|l|}{// 
{\em Say $E_s$ = set of e-asyncs in {\tt S1}.}} \\
\multicolumn{3}{|l|}{//
{\em $\lnot \exists e$ $\in$ $E_s$: {\tt S2} has dependence on e.}} \\
\multicolumn{3}{|l|}{// {\em {\tt S2} is not a barrier; {\tt S2} has no e-asyncs registered on clocks.}}
\\ \hline
\multicolumn{3}{|l|}{{\bf 4. Async-Finish Interchange}} \\
\begin{lstlisting}[style=rm]
$\mbox{\textbf{async}}$ $\mbox{\textbf{finish}}$ S1
\end{lstlisting}
& \hspace{-0.4cm}
$\Longrightarrow$
&
\begin{lstlisting}[style=rm]
$\mbox{\textbf{finish}}$ { $\mbox{\textbf{async}}$ S1}
\end{lstlisting}
\\ \hline
\multicolumn{3}{|l|}{{\bf 5. Finish-Method Pull}} \\
\begin{lstlisting}[style=rm]
def f2(){foo();}
def foo() {
  $\mbox{\textbf{finish}}$ S1;  }
\end{lstlisting}
& \hspace{-0.4cm}
$\Longrightarrow$
&
\begin{lstlisting}[style=rm]
def f2() {
   $\mbox{\textbf{finish}}$ { foo();}}
def foo() { S1 }
\end{lstlisting} \\
\multicolumn{3}{|l|}{ // {\em If finish-method has not been already applied on} {\tt foo()}.} 
\\ \hline
\end{tabular}
\caption{Mini Transformations to facilitate \newfinish{}}
\label{fig:new-finish-opt}
\end{figure}

{\em Finish-If Interchange} pulls out a {\tt finish} construct 
from the surrounding {\tt if} construct. 
A special case handling the if-then-else statement is shown below:\\%
\small
\begin{minipage}{5cm}
\begin{tightcode}
   if(cond)          
      \textbf{finish} S1
   else            \(\Longrightarrow\)
      \textbf{finish} S2         
\end{tightcode}
\end{minipage}
\begin{minipage}{5cm}
\begin{tightcode}
      v = cond
      \textbf{finish} \{
        if(v) S1
        else S2 \}
\end{tightcode}
\end{minipage} 

\noindent 
The switch-case statement is also handled similarly.
{\em Finish Expansion Upper} expands the {\tt finish} scope by pulling 
a preceding statement {\tt S1} in its scope. 
It requires that {\tt S1} does not have any e-asyncs registered on clocks. 
{\em Finish Expansion Lower} expands the scope 
of the {\tt finish} construct by pulling in a succeeding statement {\tt S1}.
It requires that (i) there is no dependence between {\tt S2} and the e-asyncs of {\tt S1}, 
(ii) {\tt S2} should not be a barrier, and (iii) {\tt S2} does not have any e-asyncs 
registered on clocks.
The {\em Async-Finish Interchange} interchanges the surrounding {\tt async} 
and the inner {\tt finish}.
In conjunction with other transformation rules, this rules helps to increase the scope of {\tt finish}.
{\em Finish-Method Pull} lifts a {\tt finish} construct from 
a method to all its possible callers (obtained by a conservative flow analysis).

The mini-transformations presented in Figure~\ref{fig:existing-optim} and~\ref{fig:new-finish-opt} 
can be categorized under two heads (a) main rules: transformations to eliminate redundant {\tt
finish} constructs, and (b) helper rules: transformations to expose opportunities
for applying main rules.
For example, Rule~\#2, and~\#3 (in Figure~\ref{fig:existing-optim}) reduce the static {\tt finish}  
operations; and Rule~\#1 (in Figure~\ref{fig:existing-optim}) and Rule
\#5 (in Figure~\ref{fig:new-finish-opt}) can reduce the dynamic {\tt
finish} operations; these rules fall in the category of main rules.
The Rules~\#1$-$\#5 in
Figure~\ref{fig:new-finish-opt} are examples of helper rules.
Note: i)~Rule \#5 is both `main' rule and a `helper' rule. 
ii)~The listed transformations can be applied in any order.

We start applying \newfinish{} on the leaf nodes of the call graph and continue  
applying \newfinish{} on their parent nodes (by avoiding the already visited nodes 
in the call graph, to take care of cycles due to recursion).
This process continues till one of the following scenarios is reached: 
(a) {\tt finish} construct has been pushed to the main method -- and 
no further processing of code is required, or
(b) {\tt finish} construct cannot be pulled out of the method,
due to dependences, and thus partial rollback takes place:
We follow a simple all or nothing strategy for expanding the 
scope of a {\tt finish}.
If the {\tt finish} construct cannot be pulled out of a method then 
the method is reverted to its original state (the state just before the \newfinish{} 
is applied on this method).
%
%

AFE is guaranteed to halt as -- (a) the number of call sites are finite, 
(b) every method is processed only once, and  
(c) no finish constructs are added to the 
recursive call sites.

\noindent{\bf Sample Transformation}:\\
We now present the working of \newfinish{} on the input 
code shown in Figure~\ref{fig:finish-rules-example}(a). 
Assume that {\tt S1}, {\tt S2}, {\tt S3}, and {\tt S4} have no e-asyncs. 
Figures~\ref{fig:finish-rules-example}(b-h) show the effect of applying 
\newfinish{} 
on the input code. 
\newfinish{} starts by applying {\em Finish Fusion} (Figure~\ref{fig:finish-rules-example}(b)),
followed by {\em Finish-If Interchange} (Figure~\ref{fig:finish-rules-example}(c)).
Then, it applies 
{\em Async-Finish Interchange} (Figure~\ref{fig:finish-rules-example}(d)),
{\em Loop-Finish Interchange} (Figure~\ref{fig:finish-rules-example}(e)) followed by 
{\em Tail Finish Elimination} (Figure~\ref{fig:finish-rules-example}(f)).
Next, it applies {\em Finish Expansion Upper} (Figure~\ref{fig:finish-rules-example}(g)) 
followed by {\em Finish Expansion Lower}
to obtain the code in Figure~\ref{fig:finish-rules-example}(h).

\begin{figure*}
\small
\begin{minipage}{3.3cm}
\begin{tightcode}
{\bf // Example Code}
S1;
finish \{
 for(i in 0..n)\{
  async \{
   if(cond) \{
    finish S2  \(\Longrightarrow\)
    finish S3  
   \} \} \} \}
S4;

        (a)
\end{tightcode}
\end{minipage}
\begin{minipage}{3.5cm}
\begin{tightcode}
{\bf // After Finish Fusion}
S1;
finish \{
 for(i in 0..n)\{
  async \{
   if(cond) \{
    \textbf{finish} \{ \(\Longrightarrow\)
     S2;       
     S3
    \} \} \} \} \}
S4
        (b)
\end{tightcode}
\end{minipage}
\begin{minipage}{3.2cm}
\begin{tightcode}
{\bf{// After Finish-If
// Interchange}}
S1;
finish \{
 for(i in 0..n)\{
  async \{
   \textbf{finish} \{  \(\Longrightarrow\)
    \textbf{if}(cond)\{ 
     S2;  S3
    \} \} \} \} \}
S4
        (c)
\end{tightcode}
\end{minipage}
\begin{minipage}{3.5cm}
\begin{tightcode}
{\bf{// After Async-Finish
// Interchange}}
S1;
finish \{
 for(i in 0..n) \{
  \textbf{finish} \{
   \textbf{async} \{     \(\Longrightarrow\)
    if(cond) \{
     S2; S3     
    \} \} \} \} \}
S4
        (d)
\end{tightcode}
\end{minipage}

\begin{minipage}{3.5cm}
\begin{tightcode}
{\bf{// After Loop-Finish
// Interchange}}
S1;
finish \{
 \textbf{finish} \{
  \textbf{for}(i in 0..n)\{
   async \{
    if(cond) \{   \(\Longrightarrow\)
     S2;  S3   
    \} \} \} \} \}
S4
        (e)
\end{tightcode}
\end{minipage}
\begin{minipage}{3.5cm}
\begin{tightcode}
{\bf{// After Tail Finish
// Elimination}}
S1;
\textbf{finish} \{
 for(i in 0..n)\{
  async \{
   if(cond) \{     
    S2;         \(\Longrightarrow\)
    S3   
   \} \} \} \}
S4
        (f)
\end{tightcode}
\end{minipage}
\begin{minipage}{3.3cm}
\begin{tightcode}
{\bf{// After Finish 
// Expansion Upper}}
\textbf{finish} \{
 \textbf{S1;}
 for(i in 0..n)\{
  async \{
    if(cond) \{     
     S2;      \(\Longrightarrow\)
     S3;
    \} \} \} \}
S4
        (g)
\end{tightcode}
\end{minipage}
\begin{minipage}{3.3cm}
\begin{tightcode}
{\bf{// After Finish 
// Expansion Lower}}
\textbf{finish} \{
 S1;
 for(i in 0..n)\{
  async \{
    if(cond) \{
     S2;  
     S3
    \} \} \}
 \textbf{S4}    \}
        (h)
\end{tightcode}
\end{minipage}
\caption{Applying \newfinish{} on a running example}
\label{fig:finish-rules-example}
\end{figure*}

%
%
%
\subsection{\dynamicChunkingFull{}}
\label{ss:dlbc}
The existing loop-chunking (LC) optimization~\cite{nandivada13-toplas} 
suffers from a drawback that it may create
tasks even when there are no idle workers at runtime.
This may lead to significant overheads (especially in case of RTP programs, 
where it is common to have many tasks created at each level of recursion).
Our proposed \dynamicChunkingFull{} (\dynamicChunking{}) addresses this
drawback through two simple, yet
effective strategies: (i) dynamic task creation 
based on the number of idle workers and load balancing among the
workers, and (iii) serial execution 
if no idle workers are available.

\subsubsection{Dynamic task creation and load balancing}
\label{s:load-balance}
One main drawback of \lc{} is that it doesn't distribute the work equally
among the available workers.
Our chunking policy aims at balancing the load through two simple techniques: 
(a) dividing the work equally among all the idle workers, and 
(b) sparing some work for the {\em current worker} (worker executing the current
task).

To highlight the unbalanced load distribution inherent 
in \lc{} consider the code shown in
Figure~\ref{fig:motivate-nqueens}. 
Figure~\ref{fig:motivate-nqueens}(b) 
shows the code after invoking \lc{} on 
Figure~\ref{fig:motivate-nqueens}(a).
In Figure~\ref{fig:motivate-nqueens}(b), consider {\tt n=12} and number of 
total workers = {\tt nChunks} = 4. 
Thus, {\tt chunkSize} is equal to {\tt 3}, and we create four tasks 
(to execute three iterations each).
Say, excluding the current worker, the other three workers are currently
idle.
In such a scenario, two of the idle workers execute one task each, and the other idle
worker will execute two tasks (six iterations),
 while the current worker waits 
at the join point for the spawned tasks to terminate.

In contrast, our chunking policy distributes the iterations equally among
all the four workers (including the current worker) -- better load balancing.
Further, the current worker gets some useful work to perform, before waiting at the join point.
Importantly, if {\tt n} = 10,
%
%
our scheme provides two iterations each to the current worker and an idle worker, and 
three iterations each to the remaining two workers.
Thus, our policy ensures that the current worker 
not only does some useful work (before waiting at the join point), but also
gets the smallest chunk of iterations to execute.


We extend the X10 Runtime (XRX) with  a function  {\tt Runtime.retIdleWorkers}()  
that returns the count of idle workers at that instant, at runtime.
Our implementation of {\tt retIdleWorkers()}does not use any atomics. 
So, in a RTP program, it is possible that two tasks may 
fetch the same value of idle workers, at the same instant. 
Thus, in  practice, the number of tasks created by \dynamicChunking{} may be 
more than the number of idle workers. 
But we show that the reduction in task creation is significant enough.
Although, the use of atomics looks lucrative, but it leads to substantial 
overheads.


Overall \dynamicChunking{} consists of five substeps (see
Figure~\ref{fig:finReductor}).
It starts by invoking LC. 
The next step is to introduce some template code that computes the current
count of the idle workers and a set of five helper variables:
i)~{\tt totWorkers}: \# idle workers+1, 
ii)~{\tt eqChunk}: minimum number of iterations executed by any worker,
iii)~{\tt actualn}: number of iterations of the parallel loop to be executed.
iv)~{\tt newN}: total number of iterations to be executed by the idle workers,
and v)~{\tt rem}: a temporary variable.
This substep also introduces an outer while loop, which is used to avoid
unstructured control flow.
The third substep of \dynamicChunking{} (Chunked-Block-Modification)
modifies the chunked code to enforce the load balancing scheme discussed
above. 
Similarly, the Parent-Block-Generation step introduces code to be executed
by the parent thread.
For the input code of Figure~\ref{fig:motivate-nqueens}(a),
Figure~\ref{fig:dynamic-chunking} shows the code generated by
\dynamicChunking{}.
The code computes the number of idle workers and 
if {\tt workers>0}, the execution continues at line~7.
The {\tt finish} body includes a chunked parallel loop
(chunked-block: executed by the idle workers), and 
a serial for-loop 
(parent-block: executed by the current worker).
\begin{figure}[t]
\small
%
%
%
\begin{tightcode}
1 def \textbf{nqueens}(val n:Int,val j:Int, ...) \{
2  var ii:Int=0;
3  var \textbf{workers}:Int = \textbf{Runtime.retIdleWorkers()};
4  
5  \textbf{outer}: while(true) \{
6     \textbf{if}(workers>0) \{ 
7        val totWorkers:Int = workers+1;
8        val actualn:Int=n-ii;
9        val eqChunk:Int=actualn/totWorkers;
10       val newN:Int=actualn-eqChunk;
11       var rem:Int=actualn\%totWorkers+workers;
12       \textbf{finish} \{
13          for( ; ii<newN; ) \{                        //\mc{\bf ``chunked block"}
14             val kx = ii+eqChunk+rem/totWorkers;
15             val ni=ii; rem--; ii = kx;
16             \textbf{async} \{
17                for(var i:int=ni ; i<kx; i++) \{
18                   ... /* {\em Checking if none of the queens conflict}  */
19                   \textbf{nqueens}(n, j, ...);
20                \} \}/* \mc{async} */\}/* \mc{outer-for} */
21          \{                                          //\mc{\bf ``parent block"}
22             for(var i:int=newN;i<size;i++)\{
23                ...    /* {\em Checking if none of the queens conflict}  */
24                \textbf{nqueens}(n, j, ...);
25             \} \} \} /* \mc{finish} */ \} /* \mc{if} */
26    \textbf{else} for(i=0; i<n; i++) \{                         //\mc{\bf ``serial block"}
27           ...         /* {\em Checking if none of the queens conflict}  */
28           \textbf{nqueens}(n, j, ...);
29           \textbf{workers} = \textbf{Runtime.retIdleWorkers()};
30           if(workers>0  \&\& i<n-2) \{
31              ii=i+1;  \textbf{continue outer};
32           \} \} break; \} /* \mc{while} */ \}  /* \mc{nqueens} */
\end{tightcode}
%
\caption{\dynamicChunking{} applied on BOTS Nqueens kernel}
\label{fig:dynamic-chunking}
\end{figure}

\subsubsection{Serial Execution}
\label{ss:serial-exec}
\dynamicChunking{} aims to create tasks only if there are idle workers. 
Ideally, if there are no idle workers then a new task should not be spawned.
In such cases the current task can be asked to complete the remaining 
job serially.
\dynamicChunking{} handles this scenario, by using a simple heuristic:
If at the time of task creation the number of idle workers are 
zero, then the loop under consideration should be executed serially.
This heuristic is enforced by invoking the Serial-Block-Generation
substep.
This substep emits sequential code to be executed when no idle workers
are found.
Considering the possibility that some workers may get freed up during 
the life-time of this serial loop, the generated code checks for available
idle workers, after each iteration.
And if idle workers are available, the rest of the iterations are divided
into {\tt totalWorkers} (= number of idle workers + 1) number of chunks to
be executed in parallel.

The ``serial block'' in Figure~\ref{fig:dynamic-chunking}, depicts the
code generated by the 'Serial-Block-Generation' substep.
An interesting point to note is that
At the end of each serial iteration, we check the 
count of the idle workers.
If that count is greater than zero (and at least two iterations are left 
to execute, to account for the work available for the current worker and at 
least one of the idle workers), we execute the remaining iterations in parallel.
To do so, 
{\tt ii} is set to  the number of iterations that have already been
executed, and
the control is transferred to line~5;
at line~8, {\tt ii} is used to compute the value of {\tt actualn}.



\subsubsection{Synchronization Operations and \dynamicChunking{} }
\label{s:dlbc-sync}
Our transformation scheme undergoes a  small tweak to handle synchronization operations, in the input code.  
Consider the input code shown in
Figure~\ref{fig:barrier-dynamic-chunking}(a) and the code generated by
\lc{} in Figure~\ref{fig:barrier-dynamic-chunking}(b).
The code generated by \dynamicChunking{} is shown in
Figure~\ref{fig:barrier-dynamic-chunking}(c).
Note that the code generated by \lc{} substep will always be of the form shown in 
Figure~\ref{fig:barrier-dynamic-chunking}(b), where the
{\tt async} body consists of a series of serial-for-loop separated by
{\tt Clock.advanceAll} statement; the serial-for-loop bounds are guarded by a condition.
%

Similar to the code shown in Figure~\ref{fig:dynamic-chunking}, the code
in 
Figure~\ref{fig:barrier-dynamic-chunking}(c) also contains three distinct
blocks ``chunked'', ``parent'', and ``serial''.
Further, the ``chunked block'' and the ``parent block'' have an additional {\tt switch} statement each.
Consider the scenario, when there are no idle workers, and the ``serial
block'' is in execution.
After executing all the iterations of {\tt S1}, we check for the availability 
of idle workers, and if available we go back to 
the ``chunked block'', to execute the instances of {\tt S2}.
The {\tt switch} statement in the ``chunked block'' helps skip the code
that is already executed in the ``serial block''.
This selection happens using the variable {\tt phase} whose value matches the
number of {\tt Clock.advanceAll} statements executed in the ``serial
block''.
We follow a similar strategy for generating code for the ``parent block''.

Note that in the ``serial block'' we do not check for the availability of
idle workers after the execution of each instance of {\tt S1}.
This is mainly done to keep a tab on the complexity of the generated code
and the overheads.

\begin{figure}
\centering
\small
\begin{tightcode}
\textbf{finish} \{ for(var i:Int=0; i<n; i++) \{
         \textbf{async clocked(c)} \{ S1; \textbf{Clock.advanceAll();} S2; \} \} \}
                      (a)
var \textbf{workers}:Int = \textbf{Runtime.retNthreads()}; 
var \textbf{chunkSize}:Int=(n+workers-1)/workers;
\textbf{finish} \{  
  for(var ii:Int=0;ii<n;ii+=chunkSize) \{
     val ni = ii;
     \textbf{async clocked(c)} \{
        var kx:Int=ni+chunkSize; if(kx>n)kx=n;
        for(var i:Int=ni; i<kx; i++) S1;
        \textbf{Clock.advanceAll()};
        for(var i:Int=ni; i<kx; i++) S2; \} \} \} 
                      (b)
var ii:Int=0, phase:Int=0;
var \textbf{workers}:Int = \textbf{Runtime.retIdleWorkers()};
\textbf{outer}: while(true) \{
   if(workers>0) \{
      val totWorkers:Int = workers+1;
      val actualn:Int = n-ii;
      val eqChunk:Int = actualn/totWorkers;
      val newN:Int = actualn-eqChunk;
      var rem:Int=actualn\%totWorkers+workers;
      \textbf{finish} \{
         for( ; ii<newN; ) \{                           {\bf //``chunked block"}              
            val kx:Int=ii+eqChunk+rem/totWorkers;
            val ni=ii; rem--; ii = kx;
            \textbf{async clocked(c)} \{
               \textbf{switch(phase)} \{
                  \textbf{case} 0:for(var i:int=ni;i<kx;i++) S1;
                        \textbf{Clock.advanceAll()};
                  \textbf{case} 1:for(var i:int=ni;i<kx;i++) S2;
               \} \} /* \mc{async} */ \} /* \mc{outer-for} */
            \textbf{switch(phase)} \{                            {\bf //``parent block"}
               \textbf{case} 0:for(var i:Int=newN;i<n;i++) S1;
                     \textbf{Clock.advanceAll()};
               \textbf{case} 1:for(var i:Int=newN;i<n;i++) S2;
            \} /*\mc{parent}*/ \} /*\mc{finish}*/ \} /*\mc{if}*/
   else /*\mc{workers <= 0}*/ \{                             {\bf //``serial block"}
      for(i=0 ; i<n; i++) S1;
      \textbf{Clock.advanceAll();}
      \textbf{workers = Runtime.retIdleWorkers()};
      if(workers>0) \{ \textbf{phase++}; \textbf{continue outer}; \}
      for(i=0;i<n;i++) S2; 
   \} /* \mc{else} */ 
   break;  \} /* \mc{while} */
                      (c)
\end{tightcode}
\caption{Synchronization operations and chunking.
(a) Unoptimized version, (b) \lc{} version, and (c) \dynamicChunking{} version.}
\label{fig:barrier-dynamic-chunking}
\end{figure}

}
 
\subsection{Possible Overheads}
\label{s:discussion}
\label{s:dlbc-overheads}

{\em Overheads due to \newfinish{}:} The code generated by \newfinish{} may incur overheads on two accounts
(i) reduction in parallelism: 
Consider the code transformation shown below: \\
\begin{tabular}{ll}
\begin{lstlisting}[style=rm]
def f2() {          
  foo(); bar(); }
def foo() {           $\Longrightarrow$       
  $\mbox{\textbf{async}}$ $\mbox{\textbf{finish}}$ S1 }
\end{lstlisting}
&
\hspace*{-0.17in}
\begin{lstlisting}[style=rm]
  def f2() {
    $\mbox{\textbf{finish}}$ { foo(); } bar(); }
  def foo() {
    $\mbox{\textbf{async}}$  S1 }
\end{lstlisting}
\end{tabular}

\noindent It can be seen that the shift of {\tt finish} construct from the method {\tt foo()} 
to its call site, inhibits the parallel execution of {\tt S1} and the
call to the function {\tt bar} (unless, the scope of the {\tt finish} can
be further expanded later to include the call to {\tt bar}). 
(ii) management of large number of (clocked) activities by a single {\tt finish}:
The task executing the join operation ({\tt finish}), performs some book keeping
such as, collecting all the exceptions, deallocating resources,
de-registering the tasks from the registered clocks (in case of clocked asyncs) and so on.
Due to its aggressive nature, AFE entrusts all these bookkeeping works of
many {\tt finish} operations (that otherwise may have run in parallel) to one
{\tt finish} operation present in a parent task.
This may lead to reduction in parallelism and performance
degradation.

{\em Overheads due to \dynamicChunking{}:} \dynamicChunking{} inserts a number of instructions to 
do load balancing, and to check for the available idle workers. 
The resulting overheads
can offset the gains, especially if these computations dominate 
the actual work done by the tasks. 
In Section~\ref{s:eval}, we show that all these overheads are compensated by
the gains resulting from \dcafe{}.
%
%
%


\section{Extensions for Exceptions}
\label{s:exception}
In this section we extend our proposed techniques to generate 
semantics preserving code in the presence of X10 exceptions
(see Section~\ref{s:back} for a brief introduction).
To motivate the impact of exceptions on the presented mini-transformations,
consider Rule~2 of Figure~\ref{fig:new-finish-opt}, being applied on the following example, where
{\tt S1} can throw an exception (of type {\tt Ex}).\\
\hspace{-0.6cm}
\begin{tabular}{ll}
\begin{lstlisting}[style=rm]
try{ S1; $\mbox{\textbf{finish}}$ S2   $\Longrightarrow$
} catch(e:Ex){...}
\end{lstlisting}
&
\begin{lstlisting}[style=rm]
try{ $\mbox{\textbf{finish}}$ { S1; S2;
} } catch(e:Ex){...}
\end{lstlisting}
\end{tabular}\\
In the LHS,
the exception thrown by {\tt S1} is caught by the {\tt catch} block.
However, in the RHS, 
the {\tt finish} block catches this exception and
  in turn
throws an object of type {\tt MultipleExceptions}.
Thus, Rule 2 is not semantics preserving, 
in the presence of exceptions. 
We now extend our transformation rules,
to address  such challenges, 

%


%
%
%

To aid the translation process, we use a temporary {\tt finish} construct 
of the form ``{\tt finish~\{S1\}$_{\mbox{\tt <exlist>}}$}'', where
{\tt exlist} represents a sequence of conditional throw statements. 
Each entry in {\tt exlist} is of the form ``{\tt if (ex != null) throw
ex;}".
We call {\tt exlist} the list of {\em pending exceptions}.
This temporary construct is translated away,
at the end of the translation process, using the following rule:
\begin{minipage}{0.7\columnwidth}
\begin{lstlisting}[style=rm]
$\mbox{\textbf{finish}}${S1}$_{\mbox{\tt <exlist>}}$   $\Longrightarrow$  $\mbox{\textbf{finish}}${S1}; exlist;
\end{lstlisting}
\end{minipage}


\subsection{\newfinish{} in the presence of exceptions}
Figures~\ref{fig:except-seman} and~\ref{fig:existing-opt} present the rules for 
doing \newfinish{} in the presence of exceptions. 
Here, we use {\tt ME} to refer to the X10 {\tt MultipleExceptions} 
class. For brevity, we avoid re-stating the old rules specified
in Figures~\ref{fig:existing-optim} and~\ref{fig:new-finish-opt} and use
``{\tt// {++}}'' to refer to the same.

\begin{figure}[t]
\centering
\small
\begin{tabular}{|p{0.4\columnwidth}p{0.48\columnwidth}|} \hline
\hspace*{-0.1in}
\begin{tabular}{l}
\multicolumn{1}{l}{\bf 1. Finish-If interchange}\\
\begin{lstlisting}[style=rm]
if(cond) {
  $\mbox{\textbf{finish}}$ {      $\Longrightarrow$
    S1}$_{\mbox{\tt <exlist>}}$}
\end{lstlisting} 
\end{tabular}
&
\begin{tabular}{l}
\begin{lstlisting}[style=rm]
v = cond;
$\mbox{\textbf{finish}}$ {
 if(cond) S1}$_{\mbox{\tt <exlist>}}$
\end{lstlisting}
\end{tabular}
\\ \hline
\hspace*{-0.1in}
\begin{tabular}{l}
\multicolumn{1}{l}{\bf 2. Finish Expansion Upper}\\
\begin{lstlisting}[style=rm]
S1;
$\mbox{\textbf{finish}}$ {        $\Longrightarrow$
  S2 }$_{\mbox{\tt <exlist>}}$ 
\end{lstlisting}\\ 
// {\em ++}  \\
// {\em e-asyncs in {\tt S1} do not}\\
// \em throw exceptions.\\
\end{tabular}
&
\hspace*{-0.1in}
\begin{tabular}{l}
\begin{lstlisting}[style=rm]
var e:Exception=null;
$\mbox{\textbf{finish}}$ { try { S1 }
  catch(e1:Exception)
   {e = e1; }
  if(e == null) S2 
}<if(e!=null)throw e;
exlist>
\end{lstlisting}
\end{tabular}
\\ \hline

\hspace*{-0.1in}
\begin{tabular}{l}
\multicolumn{1}{l}{\bf 3. Finish Expansion Lower}\\
\begin{lstlisting}[style=rm]
$\mbox{\textbf{finish}}$ {        $\Longrightarrow$
  S1 }$_{\mbox{\tt <exlist>}}$
S2
\end{lstlisting}\\
\\
// {\em ++}\\
// {\em e-asyncs in {\tt S1} and {\tt S2}}\\ 
// {\em do not throw exceptions.}
\end{tabular}
&
\hspace*{-0.1in}
\begin{tabular}{l}
\begin{lstlisting}[style=rm]
var e:Exception=null;
$\mbox{\textbf{finish}}$ { S1;
 try { exlist }
 catch(e1:Exception)
  { e = e1; }
 if(e==null){ try {S2}
  catch(ex:Exception)
   { e = ex; } } 
}<if(e!=null)throw e;>
\end{lstlisting}
\end{tabular}
\\ \hline

\hspace*{-0.1in}
\begin{tabular}{l}
\multicolumn{1}{l}{\bf 4. Async-Finish Interchange}\\
\begin{lstlisting}[style=rm]
$\mbox{\textbf{async}}$ {         $\Longrightarrow$
  $\mbox{\textbf{finish}}$ {S1}$_{\mbox{\tt <> }}$
\end{lstlisting}\\
// {\em {\tt S1} throws no exceptions.}
\end{tabular}
&
\begin{tabular}{l}
\begin{lstlisting}[style=rm]
$\mbox{\textbf{finish}}$ { 
 $\mbox{\textbf{async}}$ { S1 } 
}$_{\mbox{\tt <>}}$
\end{lstlisting}
\end{tabular}
\\ \hline

\hspace*{-0.1in}
\begin{tabular}{l}
\multicolumn{1}{l}{\bf 5. Try-Finish Exchange}\\
\begin{lstlisting}[style=rm]
try {
 $\mbox{\textbf{finish}}$ {      $\Longrightarrow$
   S1 }$_{\mbox{\tt <exlist>}}$
} catch(e:Ex)
{ S2 }
\end{lstlisting}\\ 
// {\em e-async in {\tt S1} do not} \\
// {\em throw exceptions.} \\
\end{tabular}
&
%
\hspace*{-0.1in}
\begin{tabular}{l}
\begin{lstlisting}[style=rm]
var e:Ex=null;
$\mbox{\textbf{finish}}$ {try {try {S1}
  catch(e1:Exception)
   {throw new ME(e1);}
  exlist
 }catch(e1:Ex){e=e1;}
} if (e!=null){S2}
\end{lstlisting}
\end{tabular}
\\ \hline

\end{tabular}
\caption{Rules of Figure~\ref{fig:new-finish-opt}, in the presence of exceptions.}
\label{fig:except-seman}
\end{figure}

Figure~\ref{fig:except-seman} presents the modifications to our proposed 
mini-transformations in the presence of exceptions.
The {\em Finish-If Interchange} rule is similar to the one shown in 
Figure~\ref{fig:new-finish-opt}.
%
{\em Finish Expansion Upper} requires that no exceptions are thrown by 
the e-asyncs in {\tt S1}.
The transformed code catches the exception (if any) thrown in {\tt S1} and
throws the exception outside the {\tt finish}.
The execution of {\tt S2} occurs only if {\tt S1} throws no exceptions.
Similarly, {\em Finish Expansion Lower} requires that no exceptions are thrown 
by the e-asyncs of both {\tt S1} and {\tt S2}; execution of 
{\tt S2} occurs only if {\tt S1} and {\tt exlist} throw no exceptions. 
{\em Async-Finish Interchange} requires that {\tt S1} 
does not throw exceptions. 
It also requires the {\tt finish} has no pending exceptions.
%
Besides the extensions to the rules from Figure~\ref{fig:new-finish-opt},
in the presence of exceptions, we need another transformation -- {\em
Try-Finish Exchange}. 
This transformation requires that no exceptions are thrown by e-asyncs in {\tt S1}. 
%
For the ease of explanation, we explain the modifications to the {\em Finish-Method Pull}
transformation (of Figure~\ref{fig:new-finish-opt}), 
using the following example:
\vspace{-0.1cm}
\begin{tabular}{ll}
\begin{lstlisting}[style=rm]
def bar() {
   foo(); }                  $\Longrightarrow$$~$
def foo() {
   var e:Ex;
   $\mbox{\textbf{finish}}$ S1;
   <if(e != null) throw e;> }
\end{lstlisting}
& \hspace{-0.5cm}
\begin{lstlisting}[style=rm]
 var gex:Ex;
 def bar() {
    var e:Ex;
    $\mbox{\textbf{finish}}$ { foo(); e=gex;
    }<if(e!=null)throw e;> }
 def foo() { var e:Ex; S1; gex = e; }
\end{lstlisting}
\end{tabular}\\

\noindent Here we add a new instance field {\tt gex} that will store the exception ({\tt e}) occurring 
inside the method {\tt foo()} and will throw {\tt e} at the call site of {\tt foo()}.


\begin{figure}[t]
\centering
\small
\begin{tabular}{|p{0.38\columnwidth}| p{0.52\columnwidth}|} \hline
\hspace*{-0.14in}
\begin{tabular}{l}
\multicolumn{1}{l}{\footnotesize \bf 1. Loop-Finish Interchange~ ~}\\
\begin{lstlisting}[style=rm]
for(S1;cond;S2) {
  $\mbox{\textbf{finish}}$ { S3 
  }$_{\mbox{\tt <exlist>}}$
}
\end{lstlisting}\\
// {\em ++}\\

// {\em e-asyncs in {\tt cond}, {\tt S2} }\\ 
// {\em and {\tt S3} do not throw}\\
// {\em exceptions.} \\ \\ \\ \\ \\ \\ \\ \\ \\ \\ \\ \\
\end{tabular}
&
\hspace*{-0.15in}
\begin{tabular}{l}
\begin{lstlisting}[style=rm]
S1; var e:Exception=null;
var me:ME=null,v:Boolean; 
$\mbox{\textbf{finish}}$ {
 for(; ;){ try {v=cond;}
  catch(ex:Exception)
   {e = ex; break; }
  if(e==null && v){ 
   try{S3}
   catch(ex:Exception){
    me=new ME(ex);break;}
   if(me==null) {
    try { exlist }
    catch(ex:Exception)
    { e = ex; break; }
    if(e==null){
     try{S2}
     catch(ex:Exception)
     {e=ex; break;}}}}}}
<if(e!=null) throw e;
if(me!=null) throw me;>
\end{lstlisting}
\end{tabular}
\\ \hline

\hspace*{-0.14in}
\begin{tabular}{l}
\multicolumn{1}{l}{\footnotesize\bf 2. Finish Fusion}\\
\begin{lstlisting}[style=rm]
$\mbox{\textbf{finish}}${S1}$_{\mbox{<exlist}_1>}$ 
$\mbox{\textbf{finish}}${S2}$_{\mbox{<exlist}_2>}$ 
\end{lstlisting}\\
// {\em ++}\\
// {\em e-asyncs in {\tt S1} and {\tt S2}}\\
// {\em do not throw exceptions.}
\end{tabular}
&
\hspace*{-0.15in}
\begin{tabular}{l}
\begin{lstlisting}[style=rm]
$\mbox{\textbf{finish}}$ {
  S1
  exlist$_1$
  S2
}$_{<\mbox{exlist}_2>}$
\end{lstlisting}
\end{tabular}
\\ \hline

\hspace*{-0.14in}
\begin{tabular}{l}
\multicolumn{1}{l}{\footnotesize \bf 3. Tail Finish Elimination}\\
\begin{lstlisting}[style=rm]
$\mbox{\textbf{finish}}$ {
 $\mbox{\textbf{finish}}$ { 
  S1}$_{\mbox{<exlist}_1>}$
}$_{\mbox{<exlist}_2>}$
\end{lstlisting}
\end{tabular}
&
\hspace*{-0.15in}
\begin{tabular}{l}
\begin{lstlisting}[style=rm]
try { $\mbox{\textbf{finish}}$ { S1 }  
 exlist$_1$;
} catch(e:Exception) {
 val me = new ME(e);
 throw me; }$_{<\mbox{exlist}_2>}$
\end{lstlisting}
\end{tabular}
\\ \hline
\end{tabular}
\caption{Rules of Figure~\ref{fig:existing-optim}, in the presence of
exceptions.}
\label{fig:existing-opt}
\end{figure}

Figure~\ref{fig:existing-opt} presents the extensions 
for the three mini transformations of Figure~\ref{fig:existing-optim},
in the presence of the exceptions.
Rule\#1 ensures that 
{\tt S3} is executed only if no exceptions are thrown by {\tt cond}, 
{\tt S2} and {\tt exlist}.
Rule\#2 ensures that {\tt S2} is executed only if no exception is thrown 
in exlist$_1$.
Rule\#3 uses a try-catch block 
to capture the exceptions thrown by the inner 
{\tt finish} and {\tt exlist$_1$}, and rethrow it later.


\subsection{\dynamicChunking{} in the presence of exceptions}
Note that \dynamicChunking{} invokes \lc{} as its first substep.
And \lc{} is semantics preserving in the presence of exceptions.
It can be easily seen that the code introduced by \dynamicChunking{} does not 
alter the program semantics (even in the presence of exceptions). 

\section{Evaluation}
\label{s:eval}
In this section we evaluate our proposed optimizations:
\newfinish{} and \dynamicChunking{}.
We analyze these optimizations on two different systems -- a 
16 core Intel system (2 {Intel~E5-2670 2.6GHz} processors  $\times$ 8 cores
per processor)
and a 64 core AMD system (4 AMD Abu Dhabi 6376 processors $\times$ 16
cores per processor).

We implemented \newfinish{} and \dynamicChunking{}, as whole program
optimization techniques, in the x10-2.3.0-linux compiler
%
and present an evaluation of our optimizations using the 
Native X10 (C++) backend.
Each execution time reading is reported by taking an average over ten runs.
We evaluate our optimizations on a set of eight RTP
kernels (listed in  Figure~\ref{fig:benchmark-info}), where data parallel loops are the only means of specifying parallelism.
The first  five are taken from the IMSuite~\cite{imsuite13-arxiv} and
the rest three are part of the BOTS~\cite{bots} benchmark suite.
Note that, {\em BFS}, {\em DST}, and {\em MST} also have their non-clocked
versions in IMSuite. 
But we chose the clocked versions owing to their added complexity related
to barriers.

Figure~\ref{fig:benchmark-info} (first two columns) provides a brief overview of the benchmarks 
and their respective input data sets. 
For each BOTS benchmark, we list the input type (e.g., Large, Medium) and
for each IMSuite benchmark, we list the input size and a note if we are
using the standard input or a modified one.
For all the benchmarks (except {\em DST} and {\em MST}) we have 
used one of the standard inputs provided.
For {\em DST} and {\em MST} we found that the default inputs were not leading to 
much recursion (as the diameter of the input graph was around 2 or 3), thereby 
rendering the program nearly non-recursive.
To overcome this challenge, we used their respective input generators
(provided by IMSuite) to generate 
larger and denser graphs.
In the modified inputs, we cap the maximum number of neighbors of 
any node to be at 40\% of the total nodes;
the default inputs have no such limit, thereby generate graphs with very
small diameter. 
For all the benchmarks, the chosen input size was the largest input such that 
the corresponding input program  takes not more than an hour, when run on our 16-core Intel system.

%
%

\subsection{Dynamic characteristics}
Figure~\ref{fig:benchmark-info} includes the dynamic characteristics
of the benchmarks under consideration.
We executed these kernels on the specified inputs and collected the dynamic counts 
for the task creation ({\tt async}) and task termination ({\tt finish}) operations.
The last two columns of Figure~\ref{fig:dynamic-count}, present these characteristics 
for the unoptimized (\UnOpt{}), Loop Chunking (\lc{}) and \dcafe{} 
(= \dynamicChunking{}+\newfinish{}) versions. 

It can be seen that 
in comparison to both the UpOpt and LC versions,
\dcafe{} achieves a 
significant reduction in the number of {\tt async} and {\tt finish} constructs, 
for {\em BFS}, {\em NQ} and {\em BY} kernels. 
For {\em DR}, {\em HL} and {\em FL} there is a significant reduction in the number of 
{\tt async} operations but as \newfinish{} is not able to pull out many of
the {\tt finish} 
constructs (due to MHBD), 
a substantial reduction in the number of {\tt finish} operations is not achieved. 
%
In case of {\em DST} and {\em MST} as the number of {\tt finish} and {\tt async} 
operations is low (for the UnOpt and LC versions), 
the reduction in their counts (because of \dcafe{}) is also less. 

\begin{figure}[t]
\centering
\small
{
\begin{tabular}{|l|c|c|r|r|} \hline
Kernel 		& Input  	& Type 		& \#Finish	& \#Async	\\ \hline
Breadth		& 256		& UnOpt		& 58k		& 950k		\\ \cline{3-5}
First		&       	& LC		& 31k		& 379k		\\ \cline{3-5}	
Search$^*$({\em BFS})		& (Standard) 	& DCAFE		& 1		& 64		\\ \hline
Byzantine	& 128		& UnOpt     	& 276k		& 3869k		\\ \cline{3-5}	
({\em BY})	&       	& LC        	& 276k		& 3308k		\\ \cline{3-5}	
		& (Standard)	& DCAFE		& 34		& 18k		\\ \hline
Dijkstra	& 512		& UnOpt     	& 28k		& 631k		\\ \cline{3-5}	
Routing		&       	& LC        	& 28k		& 338k		\\ \cline{3-5}	
({\em DR})	& (Standard)		& DCAFE		& 17k		& 23k		\\ \hline
Breadth		& 2048		& UnOpt     	& 3.2k		& 26k		\\ \cline{3-5}	
First		&       	& LC        	& 3.2k		& 1k		\\ \cline{3-5}	
Search$^*$({\em DST})		& (Modified)	& DCAFE		& 18		& 338		\\ \hline
Minimum		& 512		& UnOpt  	& 3.1k		& 6.3k		\\ \cline{3-5}	
Spanning 	&       	& LC        	& 3.1k		& 2k		\\ \cline{3-5}	
Tree$^*$({\em MST})	& (Modified)		& DCAFE		& 1.1k		& 1.5k		\\ \hline
Nqueens		&		& UnOpt     	& 26993k	& 377901k	\\ \cline{3-5}	
({\em NQ})	& (Large)	& LC        	& 26993k	& 377901k	\\ \cline{3-5}	
		&        	& DCAFE		& 1		& 3460k		\\ \hline
Health		&		& UnOpt     	& 17516k	& 630575k	\\ \cline{3-5}	
({\em HL})	& (Large)	& LC        	& 17516k	& 210192k	\\ \cline{3-5}	
		& 		& DCAFE		& 1636k		& 2851k		\\ \hline
Floorplan		&		& UnOpt     	& 3678k		& 19244k	\\ \cline{3-5}	
({\em FL})	& (Medium)	& LC        	& 3657k		& 19193k	\\ \cline{3-5}	
		&         	& DCAFE		& 3619k		& 1650k		\\ \hline
\end{tabular}
\caption{\small Benchmark statistics; starred(*) ones have barriers.}
\label{fig:dynamic-count}
\label{fig:benchmark-info}
}
\end{figure}

\subsection{\bf Comparing \dcafe{} Vs \lc{}}
\begin{figure*}
\begin{subfigure}{\columnwidth}
\centering
                \includegraphics[height=0.4\textwidth,width=\textwidth]{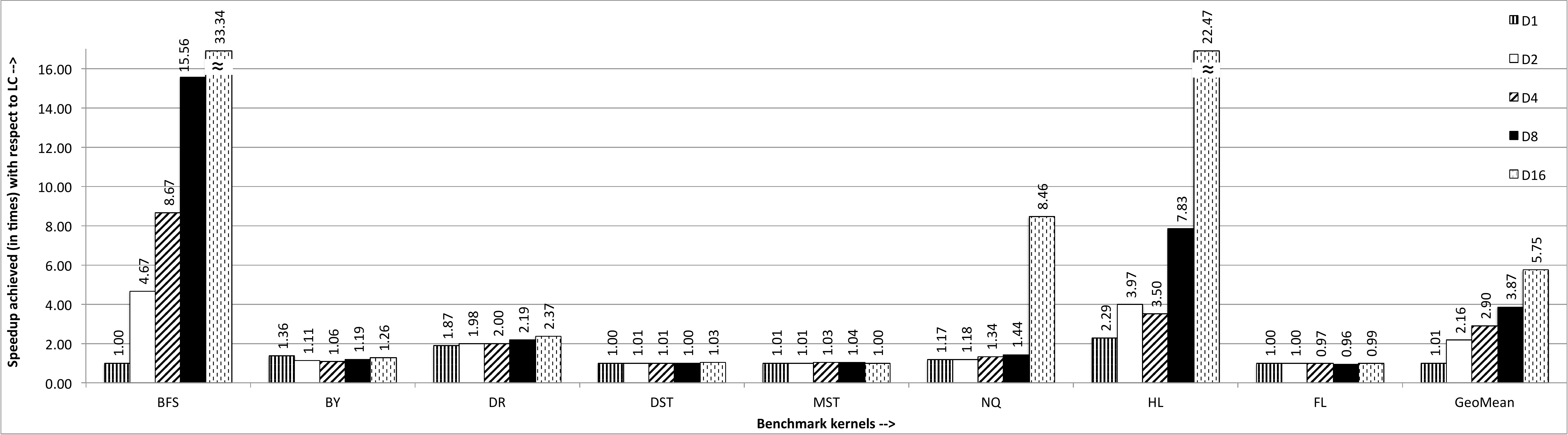}
                \caption{Intel 16-core system; varying runtime configuration D$n$, where $n$ = \#cores = 
{\tt X10\_NTHREADS.}}
		\label{fig:virgo-kernels}
\end{subfigure}

\begin{subfigure}{\textwidth}
\centering
                \includegraphics[height=0.4\textwidth,width=\textwidth]{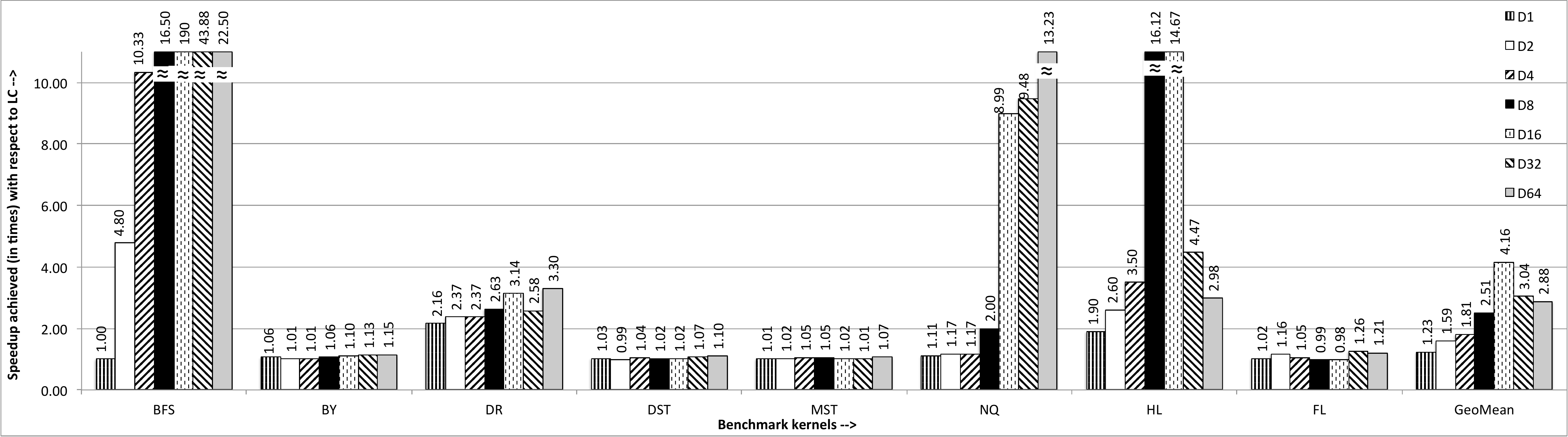}
                \caption{AMD 64-core system; varying runtime configuration D$n$, where $n$ = \#cores = 
{\tt X10\_NTHREADS.}}
		\label{fig:k2-kernels}
\end{subfigure}
\caption{\small Speedups for varying number of cores; 
Speedup~=$\frac{execution~time~of~LC~version}{execution~time~of~DCAFE~version}$}
\label{fig:perf-kernels}
\end{figure*}
Figure~\ref{fig:perf-kernels} compares the
performance of \dcafe{} with respect to \lc{}, 
for varying number of cores (in the powers of two).
%
%
Figure~\ref{fig:virgo-kernels} presents the speedups resulting from 
\dcafe{} over the \lc{} policy on the Intel system; higher the better.
We vary the number of cores and the {\tt X10\_NTHREADS} from 1 to 16, 
in sync (i.e., for simulations on a 
4 core setup, we set {\tt X10\_NTHREADS} to 4).
The performance improvement, for each kernel depends on a varied set of factors --
 the behavior of the kernel, the scope for 
reducing the task creation ({\tt async}) and the task termination 
({\tt finish}) operations, the nature of the input, runtime/OS related factors 
and the hardware characteristics.
  
It can be seen that for kernels {\em BFS}, {\em DR}, {\em NQ} 
and {\em HL}, our technique achieves significant speedups on increasing
the number of cores (and thus incrasing values of {\tt X10\_NTHREADS}). 
These speedups can be attributed to the varied effects of increased
parallelism on \lc{} and \dcafe{}.
As the number of {\tt X10\_NTHREADS} increases, \lc{}  
creates more number of tasks at each level.
In contrast, \dcafe{} creates tasks, only if idle workers are available,
and thereby is able to take advantage of the increased number of cores.
Thus, comparatively \dcafe{} has low overheads and synchronization costs, which improve its 
relative performance.
This is one of the main reasons for the sudden peak in case of {\em NQ} at 16 cores: 
the execution time for \lc{} increases sharply due to excessive task creation, while the \dcafe{} version maintains its
scalable nature (uniform decrease in execution time), 
as the number of cores are increased. 
For {\em HL} we observe a dip in its performance on moving from 2 cores to 4 cores. 
This behavior is not due to any deterioration in the performance  
of \dcafe{} version or improvement of performance of the \lc{} version for four cores, 
but because of the comparatively lower performance of the \lc{} version at
two cores.
We hypothesize this behavior of the \lc{} generated code to the system
specific scheduling policies.

A general observation is that when the number of cores are less (1 and 2) the 
performance gains for \dcafe{} are insignificant in comparison to \lc{}. 
This can be attributed to the fewer opportunities for expressing parallelism 
and the smaller value of {\tt X10\_NTHREADS}.
For such a setup, both the \dcafe{} and the \lc{} create few tasks 
at each level. 
Thus, \dcafe{} is not able to record significant task reductions and show
gains.
%

For kernels {\em DST} and {\em MST}, \dcafe{} is unable to achieve 
significant speedups over \lc{}.
This behavior can be attributed to the fewer opportunities for reduction 
of task creation and termination operations (number of {\tt async} and 
{\tt finish} operations $<3$k, see Figure~\ref{fig:dynamic-count}).

{\em FL} is an interesting kernel where, at times, \dcafe{} performs worse 
than \lc{}. 
In {\em FL} the task creation occurs inside a doubly nested loop, while the 
{\tt finish} construct is outside the nested loops.
Also, the {\tt finish} construct cannot be eliminated due to dependencies.
Importantly, the inner loop does not spawn enough tasks (to optimize to see visible
gains). 
Due to these factors, the \dcafe{} versions do not have enough scope for improvement, but do more serial work
compared to the \lc{} versions (see Section~\ref{s:dlbc-overheads}),
which in turn affects its comparative performance.



In case of {\em BY}, although \dcafe{} decreases the number of task creation and 
termination operations by a good measure, the performance gains are minimal. 
We find that {\em BY} is the only kernel where the \UnOpt{} version 
performs better than both \lc{} and \dcafe{}.
This curious behavior results from the nature of {\em BY} and the 
density of input.
In case of {\em BY}, similar to {\em FL}, there isn't much opportunity for loop chunking, 
Further, importantly, the work done by majority of the spawned tasks in 
{\em BY} is negligible. 
However, compared to the \UnOpt{} version, the \lc{} version 
introduces additional work in each task (to calculate the {\tt chunks} and so on). 
And this additional work adds to the time taken by the \lc{} versions. 
We can see that 
\dcafe{} is actually successful in bridging the gap between the performance 
of \lc{} and the \UnOpt{} to some extent.
This could be possible only due to the significant decrease in number of {\tt finish} 
and {\tt async} operations.
However, as discussed in Section~\ref{s:dlbc-overheads}, the overheads of 
\dcafe{} amortize the overall performance gains.


Figure~\ref{fig:k2-kernels} shows the behavior for the eight kernel 
benchmarks on the 64 core AMD system.
In these plots, we vary the number of cores and {\tt X10\_NTHREADS} from 
1 to 64, in sync. 
%
On increasing the cores from 1 to 16, we observe 
that the performance of the kernels is similar to that of Figure~\ref{fig:virgo-kernels}.
Except in case of HL, where the dip in performance discussed in the
context of the Intel system, is not seen here. 
Thus giving credence to the hypothesis that this may be related to some
system level scheduling issues.



On moving from 16 to 32 to 64 cores, we observe interesting characteristics.
For all the kernels (except {\em DR}, {\em DST} and {\em MST}), there is an increase 
in execution time (not shown), for all the three versions \UnOpt{}, \lc{} and \dcafe{}.
This behavior highlights that further gains in execution time cannot be 
achieved from these kernel versions (especially, for this input), by
increasing the number of cores. 
Thus, the performance gains achieved by the \dcafe{} versions (shown in
Figure~\ref{fig:k2-kernels}) are due to the 
large performance degradation of \lc{} versions, in comparison to \dcafe{}.
For example, \dcafe{} version for {\em FL}, which had a slight dip in performance 
over \lc{} (on 8 and 16 cores), achieves performance (for 32 and 64 cores), 
due to large degradation in the execution time of the \lc{} versions. 

For kernels {\em DST} and {\em MST}, as discussed earlier (for the Intel
system), the performance gains 
are not substantial due to less opportunities for exploiting parallelism. 
In case of {\em DR} kernel, we observe that the \dcafe{} version performs 
better, 
on moving from 
16 to 32 to 64 cores, but the \lc{} version does not follow this trend. 
The \lc{} version suddenly performs better for 32 cores. 
This leads to the visible dip in the speedup of \dcafe{} over \lc{}, for 32 cores. 


Overall, with respect to the \lc{} versions, the \dcafe{} versions
achieve speedup in the ranges of 
0.1$\times$ -- 33.34$\times$ (geometric mean of 5.75$\times$), on the Intel system, and 
1.07$\times$ -- 22.5$\times$, (geometric mean of 4.16$\times$), on the AMD system.

\subsection{Performance evaluation of all the proposed techniques}
\begin{figure*}
\begin{subfigure}{\columnwidth}
\centering
                \includegraphics[height=0.4\textwidth,width=\textwidth]{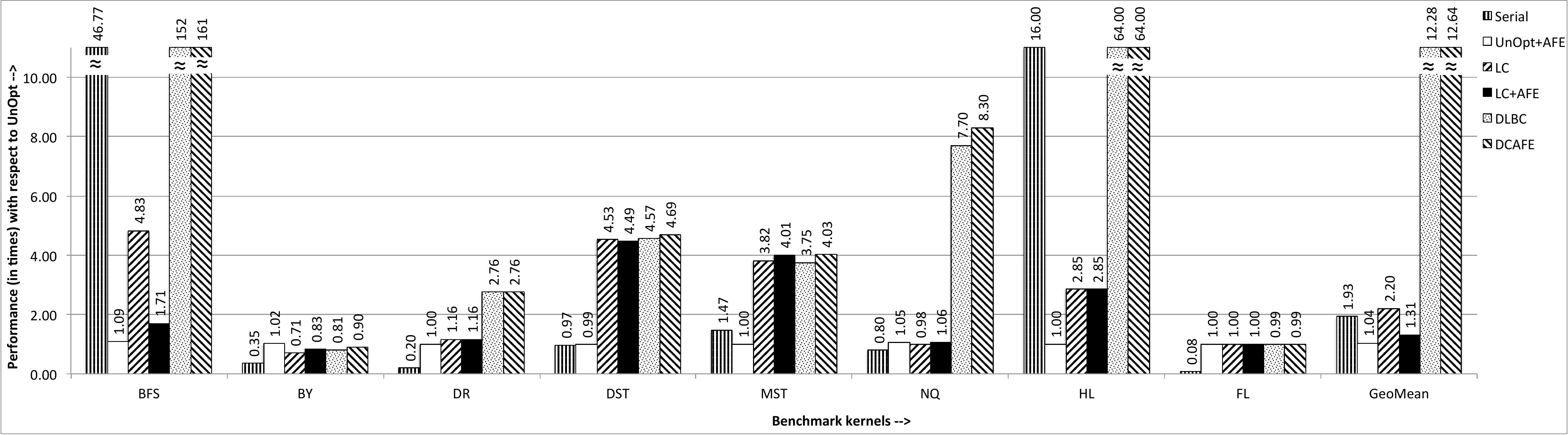}
                \caption{Intel 16-core system, configuration: \#cores = {\tt X10\_NTHREADS} = 16.}
		\label{fig:virgo-16}
\end{subfigure}

\begin{subfigure}{\columnwidth}
\centering
                \includegraphics[height=0.4\textwidth,width=\textwidth]{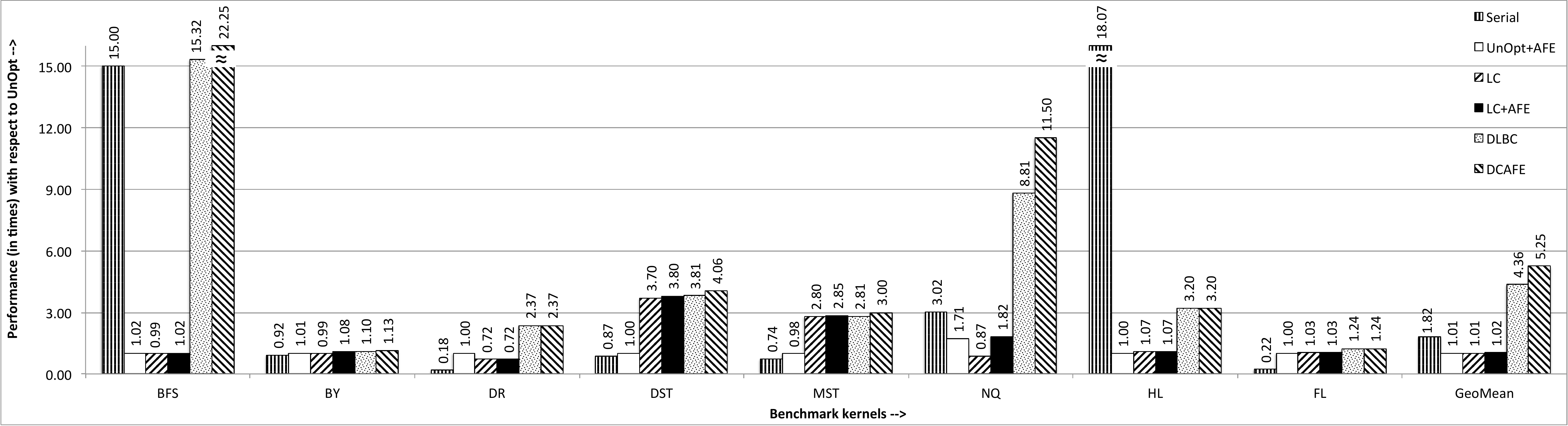}
                \caption{AMD 64-core system, static configuration: \#cores = {\tt X10\_NTHREADS} = 64.}
		\label{fig:k2-64}
\end{subfigure}
\caption{\small Comparison of different schemes with respect of UnOpt.  
{\mbox Performance of scheme $X$ with respect to UnOpt}~=~$\frac{execution~time~of~UnOpt~version}{execution~time~of~X}$}
\label{fig:perf-techniques}
\end{figure*}
We now compare the performance of Serial, \UnOpt{}+\newfinish{}, \lc, 
\lc+\newfinish, \dynamicChunking{} and \dcafe{}, with respect to
\UnOpt{}, in
Figure~\ref{fig:perf-techniques}. 
For brevity, we evaluate the kernels only for the 
largest set of hardware cores (i.e. 16 cores on Intel 
system and 64 cores on AMD) and {\tt X10\_NTHREADS} is set to \#cores.
All the results are normalized with respect to the execution times for the 
\UnOpt{} versions.

It can seen that
\newfinish{} does not reduce the number of redundant {\tt finish}
constructs for kernels {\em DR}, {\em HL} and {\em FL};
and hence \newfinish{} has no effect (as shown by the numbers of (i) \dcafe{} Vs \dynamicChunking{}, 
(ii) \lc{} Vs \lc+\newfinish{}, and (iii) \UnOpt{} Vs \UnOpt{}+\newfinish{}).
It can be seen that \dynamicChunking{} and \lc+\newfinish{} 
perform better than \lc{} (most of the time), in the context of RTP programs.
The exact performance improvement 
may differ from one kernel to another (depending on the amount of
available parallelism).
These two techniques when used in conjunction (as \dcafe), 
perform significantly better compared to all the presented techniques.
%

For {\em DST}, {\em MST} and {\em FL}, as mentioned earlier, the performance 
improvement may not be significant, and can rather have a slight dip, 
as there is limited scope for task reduction.

In case of {\em BFS}, we see a significant 
drop in the performance for \lc+\newfinish{} on the Intel system, but 
the plot for the AMD system does not show such a dip.
We ran the same benchmark on the AMD system for 16 cores and found that 
the  \lc+\newfinish{} version showed a similar behavior. 
We observed that as the number of cores increase the performance of
\lc{}+\newfinish{} version of {\em BFS} improves.

Considering the impact of \UnOpt{}+\newfinish{},
it can be seen that the \newfinish{} alone is unable to achieve much
performance difference, even in the kernels where \newfinish{} leads to
reduction in the number of \finish{} operations. 
This is due to the overheads arising out of the increased bookkeeping
activities (see Section~\ref{s:discussion}), that neutralize the gains.

For {\em BFS}, {\em HL} and {\em MST}, the \UnOpt{} versions perform worse
than the Serial version, because of 
the overheads due to
parallelization (such as cost for task creation, task termination and
barriers).
However, \dcafe{} is able to reduce these overheads and realize gains.
{\em HL} shows an interesting scenario, where \dcafe{} performs better
than Serial in Figure~\ref{fig:virgo-16}, but performs poorly in
Figure~\ref{fig:k2-64}.
On further investigation we found that the \dcafe{} version of {\em HL} actually
performs better than the serial, when it is run on 8 and 16 cores (on the
AMD system).
This is consistent with the performance of \dcafe{} shown in
Figure~\ref{fig:k2-kernels}.
A similar reason holds for 
{\em NQ}, where the Serial version performs better than the \UnOpt{} version in Figure~\ref{fig:k2-64}, but not 
in Figure~\ref{fig:virgo-16}.

Overall, it can be seen that 
compared to \dynamicChunking{},
\newfinish{} reaps less performance
improvements. 
But we argue that its impact cannot be ignored. 
Skipping the benchmarks (DR, HL and FL), where \newfinish{} did not do any
transformation, 
it can be seen that the impact of \newfinish{} is between 
1.8\% to 45.9\%, which we believe is significant.

To summarize: with respect to \UnOpt{},
our techniques \lc+\newfinish{}, \dynamicChunking{} and \dcafe{} achieve speedups 
(geometric mean)  
of 1.31$\times$, 12.28$\times$ and 12.64$\times$, respectively;
compared to these, \lc{} achieves a speedup of only 
2.2$\times$, on the Intel system.
Similarly, it can be seen that 
on the AMD system \lc+\newfinish{}, 
compared to the \UnOpt{} version,
\dynamicChunking{} and \dcafe{} achieve a speedups (geometric mean) of
1.02$\times$, 4.29$\times$ and 5.25$\times$, respectively;
compared to these \lc{} achieves a speedup of only
1.01$\times$ over \UnOpt{}.


\subsection{Energy Consumption}
We now discuss the effect of \dcafe{} and \lc{} on the energy consumption of the
benchmark kernels (on the Intel system).
%
We implemented a function {\tt read\_msr} that uses the Intel {\em Running Average Power Limit}  (RAPL)~\cite{rapl} interface to read the energy consumption of all the cores of a node.
We modify the compiler to emit a call to
this function, before 
and after the execution phase 
to calculate the energy difference.
We couldn't find a similar interface for our AMD system.

\begin{figure}[t]
	\centering
\includegraphics[width=0.6\textwidth]{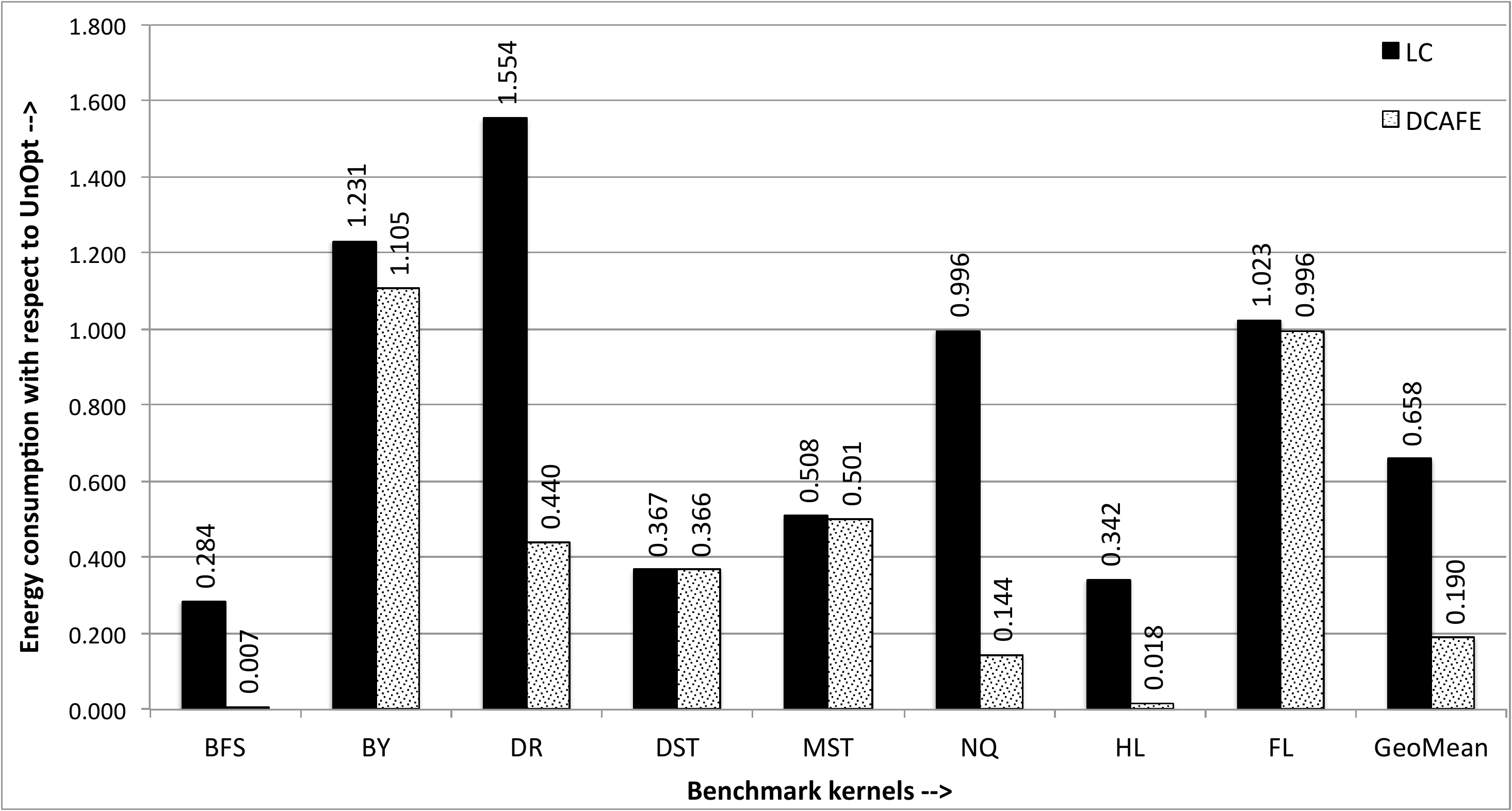}
\caption{Energy consumption normalized to \UnOpt{}.}
\label{fig:energy-hist}
\end{figure}

Figure~\ref{fig:energy-hist} depicts the energy consumed by the 
\lc{} and \dcafe{}  versions of the eight kernel benchmarks.
All the results are normalized to their 
\UnOpt{} counterparts.
It can be seen that for most of the benchmarks, both \lc{} and \dcafe{} versions
show reduction in energy consumption. 
However, the reduction due to \dcafe{} is much more than that resulting from \lc{}. 
Overall, it can be seen that compared   to the \UnOpt{} versions, the \dcafe{} versions 
consume energy in the range of 0.007$\times$ -- 1.105$\times$ (geometric mean of 0.19$\times$), 
while the \lc{} versions consume energy in the range of 0.284$\times$ -- 1.554$\times$ 
(geometric mean of 0.658$\times$). 
Overall, compared     to the \lc{} versions,
the \dcafe{} versions consume energy in the range of 0.026$\times$ 
-- 0.999$\times$ (geometric mean 0.288$\times$). 
Thus, on average the \dcafe{} versions consume 71.2\% less energy than
the \lc{} counterparts.

We observe that, maximum energy 
savings is achieved for kernels {\em BFS}, {\em DR}, {\em NQ} and {\em HL}.
These savings directly follow the significant reduction in the execution time,
which in turn is due to the reduction in
task creation and termination operations for these kernels.
On the other hand, for {\em DST}, {\em MST} and {\em FL}, there 
isn't a significant reduction in the energy consumption, which can 
be attributed to the less task reduction opportunities
available in these kernels.
In case of {\em BY}, 
compared to the \UnOpt{} version,
the energy consumption of both \dcafe{} and 
\lc{} versions is higher (follows the trend of the execution time),
However, it can be seen that \dcafe{} reduces the energy 
overheads of LC to a large extent.


\section{Discussion}
\label{s:analysis}
In this section we discuss some general discussion about our proposed
optimizations, their scope and alternatives.

~\\\noindent{\bf Non-triviality of \dynamicChunking{}}: 
To optimize RTP programs with loops using low level synchronization
primitives (like X10 clocks), \dynamicChunking{} includes many non-trivial
extensions to \lc{}.
These include i) the scheme of executing the loop serially and doing so for a
subset of iterations, before proceeding to create parallel tasks to
execute the rest of the iterations; and ii) conditionally executing the loop in parallel and
ensuring that the parent worker does some useful work, besides waiting
for the other threads to join.
These proposed extensions give rise to many interesting design choices:
i) how/when to switch between serial and parallel codes, 
ii) procedure to compute the chunking factor, 
iii) procedure to identify the idle count of worker threads and so on.
Besides the particular design choices described in Section~\ref{ss:dlbc},
we tested many other alternatives and finally zeroed in on the most
profitable ones.
Some of the choices we tested are listed below for pedagogy.
%
(a) {\em Static cut-off based on the recursion depth} -- 
This scheme stops creating new parallel tasks, once the depth of the
recursion crosses a certain static cut-off value (such as, 2, 3, 4, and
5).
Thus in this scheme, based on the cutoff, we keep creating parallel tasks even if there are no
free workers. 
Similarly, even if there are free workers, we do not create new parallel
tasks after the pre-specified recursion depth. 
Pros: Simple to implement.
Cons: Hard to predict the optimal cutoff and
minimize the overheads.
Our conclusion after experimentation: Overall inefficient and impractical.
(b) {\em Trade-offs in the serial block} -- To avoid checking for
available parallel workers after each serial iteration, we tried 
the strategy of checking for available workers only after a fixed number
of serial iterations (e.g., 2, 3, 4). 
The main intuition was to allow parallel execution when there are sufficient 
number of workers.
Pros: Reducing the overhead of checking for available workers and waiting
for sufficient number of workers.
Cons: May miss some chances to parallelize some iterations. 
Our conclusion after experimentation: The complexity of the additional checks did not pay off.
(c) {\em Minimum number of parallel tasks instead of complete
serialization} -- 
\dynamicChunking{} turns to serial code when there are no available free
workers.
We tried a scheme, where instead of executing the loop in serial, we
divided it into two chunks -- one chunk executed as part of the current
task, and the second one is executed by a new parallel task.
Pros: Chances of workers remaining free will be small.
Cons: May end up creating more tasks than required.
Our conclusion after experimentation: The cons over weighed the pros.\\

\noindent{\bf Runtime Optimizations}: 
\newfinish{} involves elaborate dependence analysis and code transformation schemes that 
are non-local in nature (even in the absence of exceptions).  
Re-casting of \newfinish{} as a runtime optimization may seem attractive, but is both 
non-trivial and can be expensive.  
Similarly, \dynamicChunking{} requires generation of serial-code from the input 
parallel code. 
This process is non-trivial, especially in the presence of deeply nested barriers such as 
clocks.
Both \newfinish{} and \dynamicChunking{} are whole program optimizations that have 
intuitive compile-time implementation and reap runtime benefits.\\

\noindent{\bf Scope of \newfinish{} and \dynamicChunking{}}: 
\newfinish{} and \dynamicChunking{} are not restricted to only X10 and can be applied to other 
task-parallel languages with similar constructs such as HJ (async/finish) and Chapel (begin/sync). 
Further, \dynamicChunking{} can also be used in other task parallel languages such as Cilk~\cite{cilkplus} and 
OpenMP~\cite{openMP}.

\section{Related Work}
\label{s:related}
There have been several works~\cite{%
cytron90-sc,heinz93-tr,tseng95-ppopp,%
ferrer09-lcpc,%
noll12-ppopp,%
nandivada13-toplas} 
that aim to reduce the overheads resulting from useless synchronization and 
join operations. 
Cytron et al. propose reduction of synchronization constructs 
by translating input fork-join code to SPMD code with reduced number of barriers.
Heinz and Philippsen perform source to source transformations to reduce the 
barrier synchronization operations in data parallel programs.
Their optimizations target the redundant synchronization operations present in the 
synchronous FORALL statements by converting them into simplified asynchronous FORALL 
statements with reduced synchronization overheads.
Tseng extends the work of Cytron et al. by using a combined
fork-join and SPMD model to reduce synchronization overheads. 
Ferrer et al. exploit the loop unrolling transformation in the 
presence of task parallel constructs.
The authors try to aggregate multiple fine-grained tasks (by unrolling loop) 
into the larger ones to achieve performance. 
Noll and Gross propose task reduction and synchronization optimizations 
for the JIT compilers.
The authors propose an optimization that allows merging of small concurrent 
tasks into a large task.
Compared to these, our optimizations eliminate redundant task
creation and termination operations in recursive task parallel programs.
Further, we present a scheme to do the transformations in a semantics
preserving manner, even in the presence of exceptions.

Yonezawa et al.~\cite{yonezawa06-ispa} aim at reducing the barrier
synchronization operations,
by generating efficient communication code for data transfer operations 
 in a distributed application.
Similarly, Bikshandi et al.~\cite{Bikshandi09} propose 
methods to efficiently execute outer-most finish operations.
Nagarajan and Gupta~\cite{NagarajanGupta10} use speculative execution to
reduce the overheads associated with barriers.
We believe that these techniques can be used in conjunction with our
proposed AFE, to further increase the performance gains.


Nicolau et al.~\cite{nicolau09-ics} propose optimizations (via code
percolation) to reduce the 
synchronization operations such as {\tt post} and {\tt wait} that are
redundant. 
In contrast, we present techniques to expand the scope of {\tt finish}
operations to reduce the number of {\tt finish} operations, especially in
the context of recursive task parallel programs.


Our work is most closely related to the work of Nandivada et
al~\cite{nandivada13-toplas}, who present a framework to reduce task creation, 
synchronization and termination operations.
They specify a set of three techniques -- finish elimination, forall coarsening and 
(static) loop chunking -- that generates efficient code for task parallel programs.
Compared to their approach, we present an approach to do efficient loop
chunking (dynamic) and aggressive finish elimination in the context of recursive
task parallel programs.

Narayanan et al.~\cite{narayanan05-iccd} use classical loop chunking to generate power 
efficient code.
Their transformation distributes equal chunks of iterations on different processors.
To the best of our knowledge, ours is the first paper that studies the
impact of reduction in task creation and termination operations on the
energy consumed. 

Loop scheduling~\cite{optimizing-kennedy01} has been one of the most popular
techniques to efficiently execute loop nests.
Some of the popular schemes of loop scheduling are static (dividing the
all the iterations equally among the declared workers), dynamic
(the iterations are divided into many small chunks and added to a work
queue and each free worker takes a chunk from this work queue to execute),
and guided (similar to dynamic, but the size of the chunks vary dynamically).
Our proposed \dynamicChunking{} method can be seen as a
specialization of loop scheduling where 
i) iterations scheduled to be executed by
the same processor are executed sequentially, 
ii) some iterations of the parallel loop may be executed sequentially,
before dividing the rest of the loop iterations among the available
workers.



There have been many works%
~\cite{suif-wilson94,adaptive-hall98,efficient-yue96} 
that computes and assigns the optimal number of
processors / workers to execute a given loop nest and parallelize the loop
accordingly.
In contrast, we use a simple scheme of chunking parallel loops based on
the number of available worker threads (number of chunks = number of
available workers).
It would be interesting to extend our proposed \dynamicChunking{} with
more sophisticated mechanisms to compute the optimal number of worker
threads.

Voss and Eigenmann~\cite{reducing-voss99} proposed an inspector-executor model that at runtime 
decides whether to execute a loop in parallel or serially. 
The main emphasis behind this scheme is that benefits of executing a loop 
in parallel may be amortized if the overheads of parallel execution are significant. 
The authors first try to run a loop in parallel and measure its execution time. 
They next compare the obtained results with the timed results of the serial 
version of the loop and decide whether to run the next versions of this 
loop in parallel or not.


There have been several prior works that control the 
parallelism based on different kinds of thresholds (all measured at
runtime).
For non RTP programs,
some of the popular threshholds are
system load~\cite{mult-kranz89,reconciling-certner08},
size of the data structures~\cite{guided-huelsbergen94,managing-aharoni92}
giving an estimation of the time the code to be parallelized may take to
execute, and
profile based estimated workload in different
iterations~\cite{efficient-prechelt02}. 
For RTP programs, 
Duran et al~\cite{bots} show the use of 
a static value of recursion depth as a cut-off for parallelization.
Similarly, dynamic cut-offs based on runtime
parameters~\cite{adaptive-duran08} have also been used for RTP programs.
Considering the difficulties in statically determining the appropriate
recursion depth, and the overheads in the dynamic approach of Duran et
al~\cite{adaptive-duran08} (requires additional monitoring threads), we
propose a scheme to determine the number of parallel tasks based on the
number of available free workers.

Our idea of task creation based on worker availability and ``serial
block'' (in \dynamicChunking{}) can be seen as a compiler based extension of lazy-binary
splitting (LBS)
scheme~\cite{lazy-tzannes10} for RTP programs and programs with
synchronization operations.
%
%
It would be interesting to evaluate the effect of \dcafe{} on an LBS based runtime
scheduler.

\section{Conclusion}
\label{s:concl} 
In this paper, we present two new optimizations \newfinish{}
(``Aggressive Finish Elimination") and \dynamicChunking{} (``\dynamicChunkingFull{}") to reduce the task
creation and termination overheads in recursive task parallel (RTP) programs.
These optimizations improve the performance, both in terms of
execution time and
energy consumption.
We implemented \dcafe{} (= \dynamicChunking{}+\newfinish{}) in the X10v2.3 compiler
and performed experiments on two different hardware
systems 
(a 16-core Intel system and a 64-core AMD system).
Compared to the loop chunking scheme  of Nandivada et al~\cite{nandivada13-toplas},
DCAFE achieved significant improvements in 
execution time (geometric mean of 5.75$\times$ and 4.16$\times$, on the Intel and
AMD system, respectively), 
and substantial reduction in the energy consumption 
(geometric mean of 71.2\% on the Intel system).
The significant improvements in execution time and reduction in energy consumption
attest to the scope of the proposed optimizations.
Though our results are shown in the context of X10,
we believe that our proposed optimizations can be applied (with similar
effect) to
other task parallel languages like OpenMP, Chapel and HJ  that admit RTP programs.
 
\newpage
\bibliographystyle{abbrvnat}
\bibliography{benchbib}

\end{document}